\documentclass[twocolumn]{aastex62}
\usepackage{amsmath}
\usepackage{chngcntr}
\usepackage{graphicx}

\begin{document}

\title{New Spatially Resolved Imaging of the SR 21 Transition Disk and Constraints on the Small-Grain Disk Geometry}

\author{S. Sallum}
\altaffiliation{Astronomy Department, University of California Santa Cruz,\\ 1156 High St., Santa Cruz, CA 95064, USA}

\author{A.J. Skemer}
\altaffiliation{Astronomy Department, University of California Santa Cruz,\\ 1156 High St., Santa Cruz, CA 95064, USA}

\author{J.A. Eisner}
\altaffiliation{Astronomy Department, University of Arizona,\\ 933 N. Cherry Ave., Tucson, AZ 85721, USA}

\author{N. van der Marel}
\altaffiliation{NRC Herzberg, 5071 West Saanich Rd.,Victoria BC\\ V9E 2E7,Canada}

\author{P.D. Sheehan}
\altaffiliation{NRAO, 520 Edgemont Rd., Charlottesville, VA 22901, USA}

\author{L.M. Close}
\altaffiliation{Astronomy Department, University of Arizona,\\ 933 N. Cherry Ave., Tucson, AZ 85721, USA}

\author{M.J. Ireland}
\altaffiliation{Research School of Astronomy and Astrophysics,\\ Australian National University, Canberra, ACT 2611, Australia}

\author{J.M. Males}
\altaffiliation{Astronomy Department, University of Arizona,\\ 933 N. Cherry Ave., Tucson, AZ 85721, USA}

\author{K.M. Morzinski}
\altaffiliation{Astronomy Department, University of Arizona,\\ 933 N. Cherry Ave., Tucson, AZ 85721, USA}

\author{V.P. Bailey}
\altaffiliation{Jet Propulsion Laboratory, California Institute of Technology,\\  Pasadena, CA 91109, USA}

\author{R. Briguglio}
\altaffiliation{INAF-Osservatorio Astrofisico di Arcetri,\\ I-50125 Firenze, Italy}

\author{A. Puglisi}
\altaffiliation{INAF-Osservatorio Astrofisico di Arcetri,\\ I-50125 Firenze, Italy}

\correspondingauthor{Steph Sallum}
\email{ssallum@ucsc.edu}

\begin{abstract}
We present new $0.6-4\mu$m imaging of the SR 21 transition disk from Keck/NIRC2 and Magellan/MagAO. 
The protoplanetary disk around SR 21 has a large ($\sim30-40$ AU) clearing first inferred from its spectral energy distribution and later detected in sub-millimeter imaging.
Both the gas and small dust grains are known to have a different morphology, with an inner truncation in CO at $\sim 7$ AU, and micron-sized dust detected within the millimeter clearing.
Previous near-infrared imaging could not distinguish between an inner dust disk with a truncation at $\sim7$ AU or one that extended to the sublimation radius.
The imaging data presented here require an inner dust disk radius of a few AU, and complex structure such as a warp or spiral.
We present a parametric warped disk model that can reproduce the observations.
Reconciling the images with the spectral energy distribution gathered from the literature suggests grain growth to $\gtrsim2-5~\mu$m within the sub-millimeter clearing.
The complex disk structure and possible grain growth can be connected to dynamical shaping by a giant-planet mass companion, a scenario supported by previous observational and theoretical studies.
\\

\end{abstract}

\section{Introduction}
Protoplanetary disks - the circumstellar disks of dust and gas that remain after star formation - are understood to be the sites of planet formation.
High-resolution images of protoplanetary disks in the visible, infrared, and millimeter have informed our understanding of this process by revealing dust and gas properties \citep[e.g.][]{2016ApJ...828...46A}, constraining the timescale for disk dispersal \citep[$\sim$ few Myr; e.g.][]{2001ApJ...553L.153H}, and suggesting the dynamical influence of young planets \citep[e.g.][]{2015A&A...584L...4P,2018ApJ...866..110D}.
Direct images of young and forming planets will evince the details of how planets form, constraining accretion rates \citep[e.g.][]{2015ApJ...803L...4E,2015ApJ...799...16Z}, initial entropies \citep[e.g.][]{2012ApJ...745..174S}, and early atmospheric properties \citep[e.g.][]{2008ApJ...683.1104F}.
Recent direct images of protoplanet candidates both in the infrared and in accretion tracers such as H$\alpha$ have allowed us to estimate their masses and accretion rates \citep[e.g.][]{2018A&A...617A..44K,2018ApJ...863L...8W,2015Natur.527..342S}, as well as atmospheric properties \citep[e.g.][]{2018A&A...617L...2M}. 

Transition disks, protoplanetary disks with large clearings in dust, are particularly good targets for protoplanet searches.
Their solar-system-sized clearings were first identified through their spectral energy distributions \citep{1989AJ.....97.1451S} and were later confirmed in sub-millimeter imaging \citep[e.g.][]{2011ApJ...732...42A}.
Transition disk holes, gaps, and low stellar accretion rates relative to full disks \citep[e.g.][]{2007MNRAS.378..369N} can be explained by planets accreting material that would have otherwise fallen onto the star \citep[e.g.][]{1999ApJ...514..344B}.
Other disk features observed in some of these objects -- such as warps, asymmetries, and spirals seen in both the millimeter \citep[e.g.][]{2014ApJ...783L..13P,2013ApJ...775...30I,2013Sci...340.1199V} and in infrared scattered light \citep[e.g.][]{2012ApJ...748L..22M,2015ApJ...798L..44M} -- may also suggest the gravitational influence of protoplanets.
Directly detecting protoplanets in transition disk clearings would inform our understanding of disk dispersal and disk-planet interactions.

Young and forming planets are predicted to have relatively low contrast at infrared wavelengths \citep[e.g.][]{2015ApJ...803L...4E,2015ApJ...799...16Z}.
However, the large distances to star forming regions and transition disks \citep[e.g.][]{2007ApJ...671.1813T} make imaging solar system scales in the infrared extremely difficult.
The technique of non-redundant masking \citep[NRM; e.g.][]{2000PASP..112..555T} is well suited for resolving transition disk clearings at these wavelengths.
NRM turns a filled-aperture into an interferometric array, providing a smaller dark fringe,\footnote{$\lambda$/2B, where B is the longest baseline in the mask, compared to 1.22 $\lambda$/D for a filled-aperture with diameter D} and superior point spread function characterization compared to a conventional telescope.
NRM can thus probe tighter angular separations than filled-aperture imaging systems, even those equipped with high performance coronagraphs \citep[e.g.][]{2014ApJ...780..171G}.
Masking has led to the detections of both companions and circumstellar disks at separations close to or within the diffraction limit \citep{2008ApJ...678L..59I,2012ApJ...753L..38B,2012ApJ...745....5K,2015Natur.527..342S,2011A&A...528L...7H,2017ApJ...844...22S,2019A&A...621A...7W}.

We present infrared NRM and visible filled-aperture observations of the transition disk around SR 21.\footnote{SR 21 has been previously classified as a member of a binary system \citep{2003ApJ...591.1064B}, and is catalogued as SR 21A. However the two components were later shown to be non-coeval \citep{2003ApJ...584..853P} and to have different proper motions \citep{2010AJ....139.2440R}.}
SR 21 is a G3 star in Ophiuchus, at a distance of 138 pc \citep{2016A&A...595A...1G}.\footnote{Previous studies have adopted other distances for SR 21; we adjust these to a distance of 138 pc when appropriate.}
Estimates of its stellar mass range from $1$ M$_\odot$, based on millimeter observations and spectro-astrometry \citep[e.g.][]{2008ApJ...684.1323P,2009ApJ...704..496B}, to $2.5$ M$_\odot$, based on stellar evolutionary models \citep[e.g.][]{2003ApJ...584..853P}. 
It is classified as a Class II young stellar object, with an accretion rate of $10^{-7.9}~\mathrm{M_\odot~yr{^{-1}}}$ \citep{2014A&A...568A..18M}.
Its nature as a transition disk was established by \citet{2007ApJ...664L.107B}, who required an inner dust disk from 0.22--0.39 AU and a dust disk gap from 0.39--16 AU to explain the spectral energy distribution.

Followup millimeter observations have confirmed this dust clearing and further constrained the morphology of the dust disk.
A cavity extending to $\sim28$ AU was detected in 880 $\mu$m SMA data \citep{2009ApJ...704..496B}. 
Re-analysis of these observations yielded a cavity radius of $\sim40$ AU \citep{2011ApJ...732...42A}.
ALMA observations at 450 $\mu$m and 870 $\mu$m led to cavity radii estimates of 34 AU and 41 AU, respectively \citep{2015A&A...584A..16P}.
Furthermore, 450 $\mu$m ALMA data show a large-scale disk asymmetry that can be fit by a vortex, and residuals of the fits may suggest spiral structure \citep{2014ApJ...783L..13P}.

CO observations of SR21 reveal a different morphology in the gas.
Spectroastrometry from VLT/CRIRES suggest a gas disk truncation at $\sim7$ AU \citep[][]{2008ApJ...684.1323P} that may be caused by a companion.
A gas density drop within $\sim25$ AU has more recently been derived using ALMA $^{13}$CO and C$^{18}$O observations at $\sim335$ GHz \citep{2016A&A...585A..58V}.
The gas disk truncation and density drop within the millimeter dust clearing are both consistent with predictions for sculpting by planetary mass companions.

Infrared observations show small-grain material within the millimeter clearing and also suggest substellar companions.
Imaging at 8.8 $\mu$m and 11.6 $\mu$m yield characteristic mid-IR sizes of 67 and 92 mas, respectively, corresponding to 9 and 13 AU at 138 pc \citep{2009ApJ...698L.169E}.
These data suggest the presence of a compact, warm companion - possibly circumplanetary material - at a disk gap edge at approximately 6 AU \citep{2009ApJ...698L.169E}.
H band images also evince small grains within the millimeter clearing, but cannot constrain the location of the disk rim \citep{2013ApJ...767...10F}.
The best fit disk position angle appears to change with stellocentric radius, suggesting a warp or spiral structure \citep{2013ApJ...767...10F}.
Compact, asymmetric mid-IR emission, dust segregation, and disk warps may all be signs of a substellar companion in the SR 21 transition disk.

The near-infrared and visible light observations of SR 21 presented here probe tighter separations than previous direct imaging studies.
They resolve the location of the companion posited in \citet{2009ApJ...698L.169E} at wavelengths where it is predicted to be relatively bright.
In Sections \ref{sec:obs} and \ref{sec:red} we describe the observations and data reduction, respectively.
In Section \ref{sec:ana} we describe our analysis of the imaging data and spectral energy distribution, and in Sections \ref{sec:disc} and \ref{sec:conc} we present our conclusions.

\section{Observations}\label{sec:obs}

\begin{deluxetable*}{lcccccc}
\caption{Target List\label{tab:targs}}
\tablehead{\colhead{Target} & \colhead{RA} & \colhead{DEC} & \colhead{G}\tablenotemark{a} & \colhead{K} & \colhead{$f_{3.35}$}\tablenotemark{b} & \colhead{Object Type} \\
\colhead{} & \colhead{(hh mm ss.ss)} & \colhead{($\pm$dd mm ss.ss)} & \colhead{(mag)} & \colhead{(mag)} & \colhead{(Jy)} & \colhead{} 
}
\startdata
SR 21 & 16 27 10.28 & -24 19 12.62 & 12.7 & 6.7 & 1.11 & Science \\
HD 149201 & 16 34 12.39 &-25 00 04.00 & 7.8 & 5.1 & 2.85 & PSF Calibrator \\
HD 149446 & 16 35 52.37 & -25 22 53.62 & 8.3 & 5.5 & 2.01 & PSF Calibrator \\
HD 141534 & 15 50 26.59 & -22 52 20.06 & 8.7 & 6.0 & 1.17 & PSF Calibrator \\
HD 150668 & 16 43 16.36 & -18 27 29.87 & 8.4 & 5.1 & 3.01 & PSF Calibrator \\
2MASS J16223249-2500322 & 16 22 32.50 & -25 00 32.42 & 12.1 & 6.7 & 0.740 & PSF Calibrator \\
2MASS J16245663-2317027 &16 24 56.64 & -23 17 02.55 & 11.4 & 6.0 & 1.40 & PSF Calibrator
\enddata
\tablenotetext{a}{\emph{Gaia} magnitudes ($0.623~\mu$m), listed since the wavefront sensors operate at approximately R band (0.658 $\mu$m).}
\tablenotetext{b}{\emph{WISE} 3.35 $\mu$m fluxes.}
\end{deluxetable*}

\begin{deluxetable}{lcccc}
\caption{Summary of Observations\label{tab:obs}}
\tablehead{\colhead{Target} & \colhead{$\mathrm{n_{visits}}$} & \colhead{$\mathrm{n_{frames}}$} & \colhead{$\mathrm{t_{exp}}$} & \colhead{$\Delta PA$}}
\startdata
\sidehead{\textbf{April 5, 2013 - Magellan - L$'$}}
SR 21 & 10 & 70 & 1.5 & 105.88\\
HD149201 & 4 & 100 & 1& 4.54\\
HD149446 & 8 & 100 & 1.2 & 85.40\\
\sidehead{\textbf{April 10, 2017 - Keck - L$'$}}
SR 21 & 8 & 20 & 20 & 71.16\\
HD150668 & 3 & 20 & 20 & 40.84\\
HD141534 & 3 & 20 & 20 & 51.28\\
\sidehead{\textbf{June 6, 2017 - Keck - L$'$}}
SR 21 & 11 & 14 & 20 & 58.25\\
2MASS J16223249-2500322 & 6 & 14 & 20 & 56.33\\
2MASS J16245663-2317027 & 5 & 14 & 20 & 45.98\\
\sidehead{\textbf{June 7, 2017 - Keck - L$'$}}
SR 21 & 9 & 14 & 20 & 58.63\\
2MASS J16223249-2500322 & 5 & 14 & 20 & 41.87\\
2MASS J16245663-2317027 & 4 & 14 & 20 & 44.19\\
\sidehead{\textbf{April 27, 2018 - Magellan - H$\alpha$}}
SR 21 & 1 & 157 & 30 & 22.89 \\
\sidehead{\textbf{June 28, 2018 - Keck - L$'$}}
SR 21 & 5 & 20 & 20 & 25.94\\
2MASS J16223249-2500322 & 2 & 20 & 20 & 12.53\\
2MASS J16245663-2317027 & 2 & 20 & 20 & 13.93 \\
\sidehead{\textbf{June 29, 2018 - Keck - L$'$}}
SR 21 & 6 & 20 & 20 & 33.84 \\
2MASS J16223249-2500322 & 3 & 20 & 20 & 26.12 \\
2MASS J16245663-2317027 & 3 & 20 & 20 & 27.63 \\
\sidehead{\textbf{July 24, 2018 - Keck - K$\mathrm{_s}$}}
SR 21 & 8 & 20 & 20 & 44.28 \\
2MASS J16223249-2500322 & 3 & 20 & 20 & 28.87 \\
2MASS J16245663-2317027 & 3 & 20 & 20 & 28.90 \\
\enddata
\end{deluxetable}

\subsection{H$\alpha$ Observations}
We observed SR 21 at H$\alpha$ using Magellan / MagAO / VisAO \citep[][]{2013ApJ...774...94C,2015ApJ...815..108M} in spectral differential imaging (SDI+) mode \citep[e.g.][]{2006SPIE.6272E..2DB,2018SPIE10703E..0LC}.
VisAO's SDI+ mode uses a Wollaston beamsplitter to simultaneously image in two narrowband ($\Delta \lambda = 6~\mathrm{nm}$) filters centered on H$\alpha$ (0.656 $\mu$m) and the nearby continuum (0.642 $\mu$m).
The continuum channel provides a reference PSF and allows for identification of sources that have excess H$\alpha$ emission.

We collected these data in pupil-stabilized mode, which allows the sky to rotate relative to the detector throughout the night.
This facilitates calibration by keeping quasi-static speckles fixed on the detector, while true astrophysical signals rotate with the sky \citep[e.g.][]{2006ApJ...641..556M}.
Since SDI provides its own reference PSF, this dataset consists of a single visit to SR 21 that we also self-calibrate using angular differential imaging \citep[ADI; e.g. ][]{2006ApJ...641..556M}. 
While we were allocated a half night for these observations, clouds limited the total observing time to $\sim 90$ minutes, and the data were of low quality (FWHM $\sim$ 160 mas).

\subsection{Ks and L$'$ Observations}
We observed SR 21 at K$\mathrm{_s}$ ($2.2~\mu$m) and L$'$ ($3.8~\mu$m) using the 6-hole NRM at Magellan / MagAO / Clio2 \citep[e.g.][]{Close:2012,2014SPIE.9148E..04M} and the 9-hole NRM at Keck / NIRC2 \citep[e.g.][]{2000SPIE.4006..491T}.
The MagAO observations were taken in 2013, and the Keck datasets were taken between 2017 and 2018.
The effective field of view of NRM observations is $\lambda$ / $d_{hole}$, where $d_{hole}$ is the diameter of a single sub-aperture in the mask. 
This corresponds to 950 mas, 700 mas, and 400 mas for MagAO L$'$, Keck L$'$, and Keck K$\mathrm{_s}$, respectively.

All data were obtained in pupil-stabilized mode, with the sky rotating on the detector throughout the night.
We broke our observations up into ``visits", each of which was a single pointing composed of $\mathrm{n_{frame}}$ frames.
For the MagAO datasets, we positioned the first half of each visit on the top of the frame, and the second half on the bottom of the frame; for Keck the dither positions were in the top left and bottom right quadrants of NIRC2.

In addition to SR 21 we observed unresolved stars as PSF calibrators (see Tables \ref{tab:targs} and \ref{tab:obs}), following a pointing pattern: ...cal 1 - target - cal 2 - target....
To limit calibration errors due to differential refraction, we chose calibrators that were close to SR 21 on the sky.
We in general also chose calibrators with similar brightness in the wavefront sensing bandpass (R band), to ensure similar adaptive optics performance.
However, the April 2013 and April 2017 observations used calibrators that were much brighter than SR 21 at R.
This resulted in comparable calibration for the phase observables, but degraded the squared visibility calibration compared to a well-matched R-band calibrator.
For the remainder of the observations, we used calibrators that were better matched to SR 21 in the visible.

We initially select our calibrators using JMMC's \texttt{searchcal} \citep{2016A&A...589A.112C}.
This software uses an interferometrically-measured relation between photometry and stellar diameter to estimate angular diameters \citep{2004sf2a.conf..181D}.
All calibrators have small angular diameters relative to Keck's 10-meter maximum baseline.
After observing, we vet all calibrators both by checking that their squared visibilities are consistent with 1, and by fitting companion models and reconstructing images. 
No resolved structure was found for any PSF calibrator.

\section{Data Reduction}\label{sec:red}

\subsection{H$\alpha$ Imaging}

To reduce the VisAO data, we bias and dark subtract all frames before dividing by a flat field and subframing down to a field of view of 1.6$"$ (200 pixels).
We then use \texttt{pyKLIP} \citep{2015ascl.soft06001W}, a \texttt{python} implementation of Karhunen-Lo{\`e}ve Image Processing \citep[KLIP; e.g.][]{Soummer:2012} to carry out PSF subtraction. 
Since accreting companions would have lower contrast at H$\alpha$ compared to the continuum, while forward scattered light would have equal contrast in the two narrowband filters, we process the VisAO observations using KLIP in two different ways. 
First, to look for excess H$\alpha$ emission beyond that expected for forward scattering, we subtract the continuum images from the H$\alpha$ images after scaling by the ratio of the stellar H$\alpha$ to continuum flux.
The scaling step accounts for the fact that forward scattering signals would have higher flux if the stellar H$\alpha$ flux were higher.
Second, to search for scattered light signals, we add the two narrowband images together.

KLIP uses several parameters to control how the reference PSF library is built.
These include $n_{an}$, the number of annuli in the field of view that are treated independently; $n_{sub}$, the number of subsections into which each annulus is divided, $IWA$, the radius inside of which pixels are discarded; $\phi_{mov}$, the number of degrees by which the image must have rotated to be included in the library; and $n_b$, the number of Karhunen-Lo{\`e}ve basis vectors subtracted from the science image.
Different source morphologies (e.g. circumstellar disks versus planets) will be better preserved for different choices of these parameters.
For example, point-like companions are more easily detected with aggressive parameter choices (e.g. large $n_{an}$, $n_{sub}$, $n_b$ and small $\phi_{mov}$), while extended sources such as circumstellar disks may appear as a collection of point sources in an aggressive reduction.

We reduce the continuum-subtracted (SDI) observations using aggressive parameter choices ($n_{an} = 6$, $n_{sub} = 4$, $\phi_{mov} = 6^\circ$, $n_b > 10$), and the summed observations using less aggressive parameters optimized for disk imaging ($n_{an} = 1$, $n_{sub} = 1$, $\phi_{mov} = 10^\circ$, $n_b < 10$).
The aggressive number of annuli was chosen so that each annulus had a width roughly equal to the stellar FWHM.
For both reductions, we vary the $IWA$ and $n_b$; changing $IWA$ and $n_b$ does not change the resulting images significantly.
We also mask an annulus between 27 and 42 pixels (216 - 336 mas) in all raw images, since this corresponds to MagAO's control radius, where there are short lived speckles that are not easily subtracted \citep[e.g][]{2017AJ....153..264F}.
We smooth the final KLIP-processed image by a Gaussian kernel with the stellar FWHM ($\sim17$ pixels).

\begin{figure*}[ht]
\begin{center}
\begin{tabular}{c} 
\includegraphics[width=0.8\textwidth]{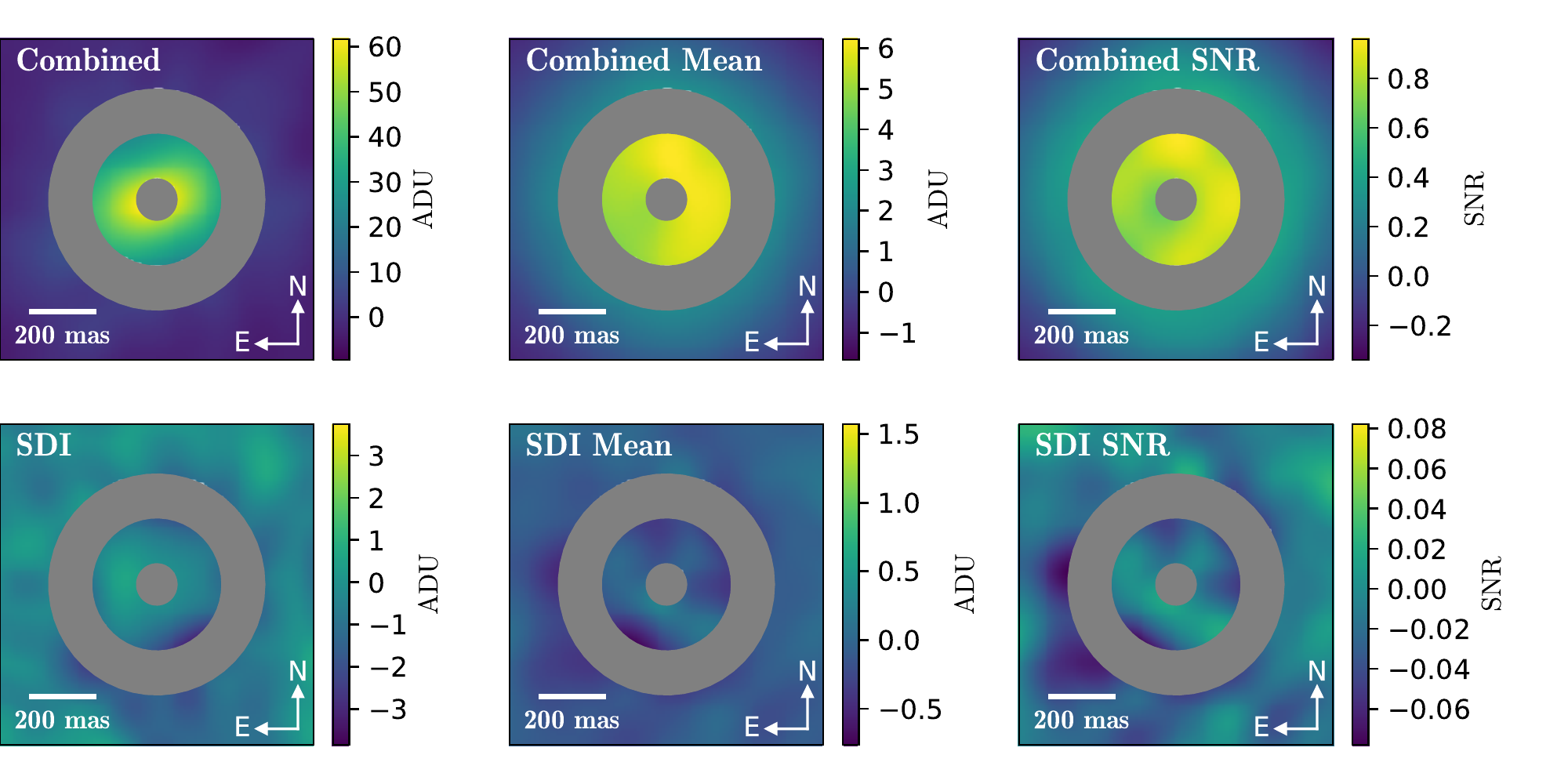}
\end{tabular}
\end{center}
\caption
{ \label{fig:klipvis}
MagAO H$\alpha$ + continuum (top), and SDI (bottom) reductions. The left panels show KLIP images for each dataset, the middle panels show mean bootstrapped images for each dataset, and the right panels show SNR maps created from the bootstrapping tests. The grey shading shows regions that have been masked in the reduction: radii less than 7 pixels, and an annulus from 27-42 pixels corresponding to the AO control radius.
}
\end{figure*}

Since the conditions and PSF quality for this dataset were variable, we also used bootstrapping to estimate mean images and SNR maps for the combined and SDI KLIP reductions.
For each reduction, we KLIP processed a large number ($>1000$) of datacubes with the same length as the MagAO data, sampling randomly from the entire set of images.
We took the mean of the set of KLIP-processed bootstrapped datacubes as a representative image for each reduction, and we used the scatter over the set to generate the SNR map. 
Figure \ref{fig:klipvis} shows the final KLIP images, the bootstrapped mean images, and the bootstrapped SNR maps.
These tests suggest that both datasets are consistent with noise. 
Since bootstrapping ADI observations effectively decreases the amount of parallactic angle coverage, we checked that we would still have detected injected companions in the bootstrapped images for both reductions.\\\\

\subsection{$\mathrm{K_s}$ and L$'$ Imaging}
We use an updated version of the data reduction pipeline described in \citet{2017ApJS..233....9S} to process the NRM observations.
For each visit, we flatten all frames and then perform dark and sky subtraction by subtracting the median of one dither position from each frame in the other dither position.
We Fourier transform the calibrated images to calculate complex visibilities that contain the amplitudes and phases corresponding to each baseline in the mask.
Due to the observing bandpass and the finite mask hole size, information from each baseline is encoded in an extended region in the Fourier transform.
To use information from the entirety of this region, we calculate closure phases using the method described in \citet{1999PhDT........19M}, in which we average bispectra for multiple triangles of pixels for each triangle of baselines in the mask.
We then average the bispectra over the cube of images in each visit before taking the phase as the mean closure phase.
Since closure phases are correlated, we project them into linearly-independent combinations of phases called kernel phases \citep[e.g.][]{2010ApJ...724..464M,2015ApJ...801...85S}.
We also calculate squared visibilities by summing the power in all pixels corresponding to each mask baseline and averaging over the cube of images.

We calibrate the kernel phases and squared visibilities using observations of unresolved stars.
We use a method presented in \citet{2017ApJS..233....9S}, in which we fit a polynomial function in time to each calibrator squared visibility and kernel phase.
We sample this function at the time of each target observation to assign instrumental kernel phases / visibilities.
We then subtract the instrumental kernel phases from the target kernel phases and divide the instrumental visibilities into the target visibilities.
We do this using $\mathrm{0^{th} - 5^{th}}$ order polynomial functions and use the order that provides the lowest kernel phase / visibility scatter to calculate the final calibrated observations.

\begin{figure*}[ht]
\begin{center}
\begin{tabular}{c} 
\includegraphics[width=0.8\textwidth]{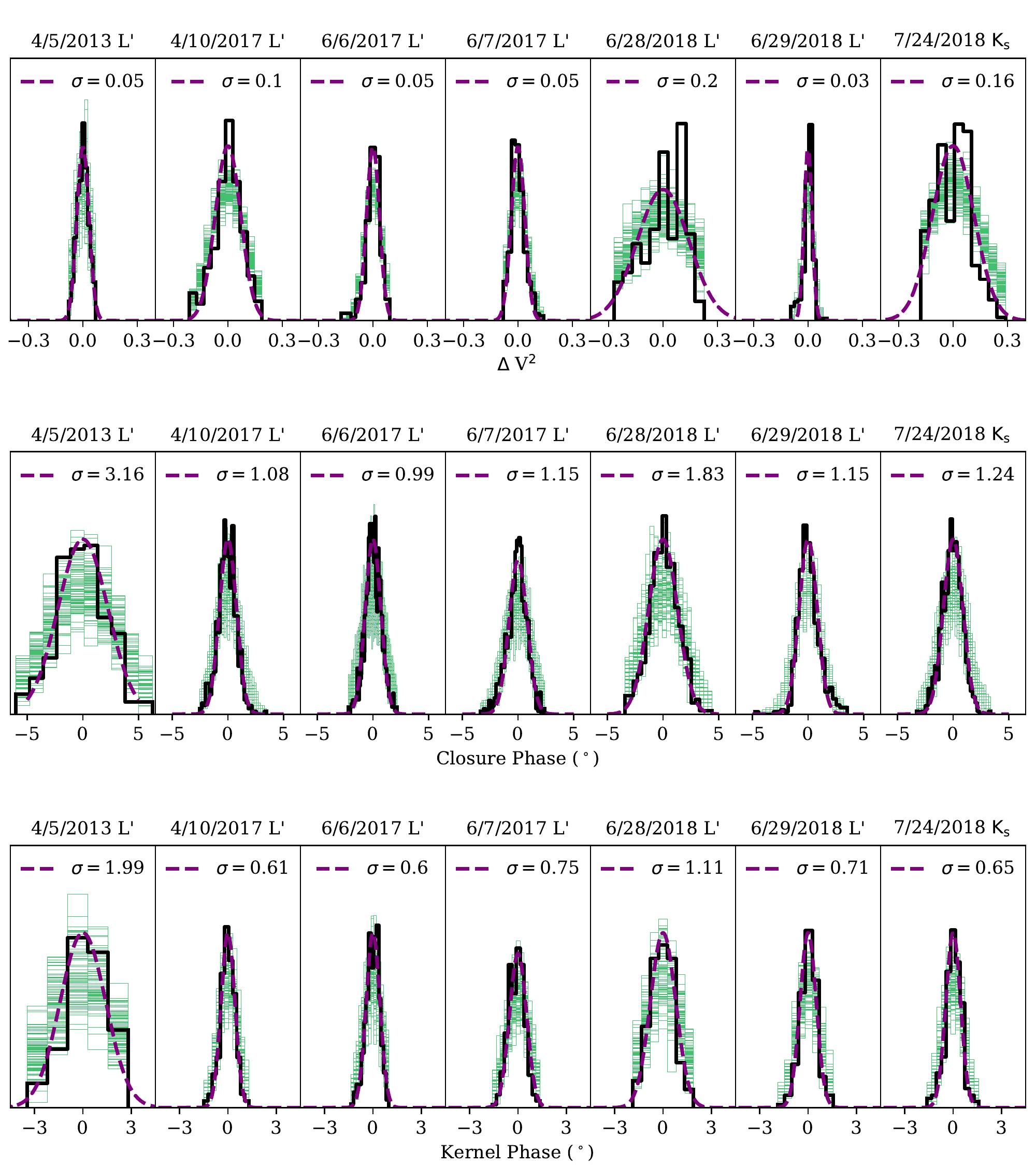}
\end{tabular}
\end{center}
\caption
{ \label{fig:hists}
Histograms for squared visibilities (top), closure phases (middle), and kernel phases (bottom) for each night of NRM observations, after subtracting the mean for that night. Black lines show the data, purple dashed lines show Gaussian fits to the black distributions, and green thin lines show histograms for simulated datasets drawn from the purple distributions.}
\end{figure*}

All of the NRM datasets are limited by calibration errors; the observed scatter in the calibrated phases and visibilities is larger than the random scatter over each cube of images.
To assign more realistic error bars, we assume that the errors for each dataset are uniform and we fit Gaussian distributions to each night of calibrated kernel phases, closure phases, and squared visibilities. 
We first subtract the mean signal for the night, to ensure that the best fit distributions have zero mean.
Figure \ref{fig:hists} compares the best fits to the observed, mean-subtracted phases and visibilities; most of the observations are well-fit by Gaussian functions.
However, the noisiest datasets are not and as a result they may have poorly-estimated errors (e.g. squared visibilities for June 28, 2018 and July 24, 2018, and closure / kernel phases for April 5, 2013). 
While the assigned errors for these datasets should thus be treated with caution, they still significantly down-weight the noisiest nights during fitting.

\section{Analysis}\label{sec:ana}
Here we present $\mathrm{K_s}$ and L$'$ images of SR 21 and use them to constrain the morphology of material within the millimeter clearing. 
We reconstruct these images from the observed closure phases and squared visibilities.
Since image reconstruction often relies on assumptions about the source morphology, we first fit parametric models to the data.
We fit simple, single companion models, which reveal asymmetries in single-epoch observations and can constrain orbital motion over multi-epoch datasets.
We also fit geometric and radiative transfer disk models to the data, which can account for static asymmetries and resolved ($< 1$) squared visibility signals (which imply extended structure).
We then reconstruct images from the observations in a parametric way, including model components that have been constrained by these simpler tests.
Due to their low quality and small amount of sky rotation, we do not include the H$\alpha$ observations in the companion and disk fitting that follows.
However, we do check that our final constraints on SR 21's morphology are consistent with the MagAO data.

\subsection{Single Companion Models}
\begin{figure*}[ht]
\begin{center}
\begin{tabular}{c} 
\includegraphics[width=0.8\textwidth]{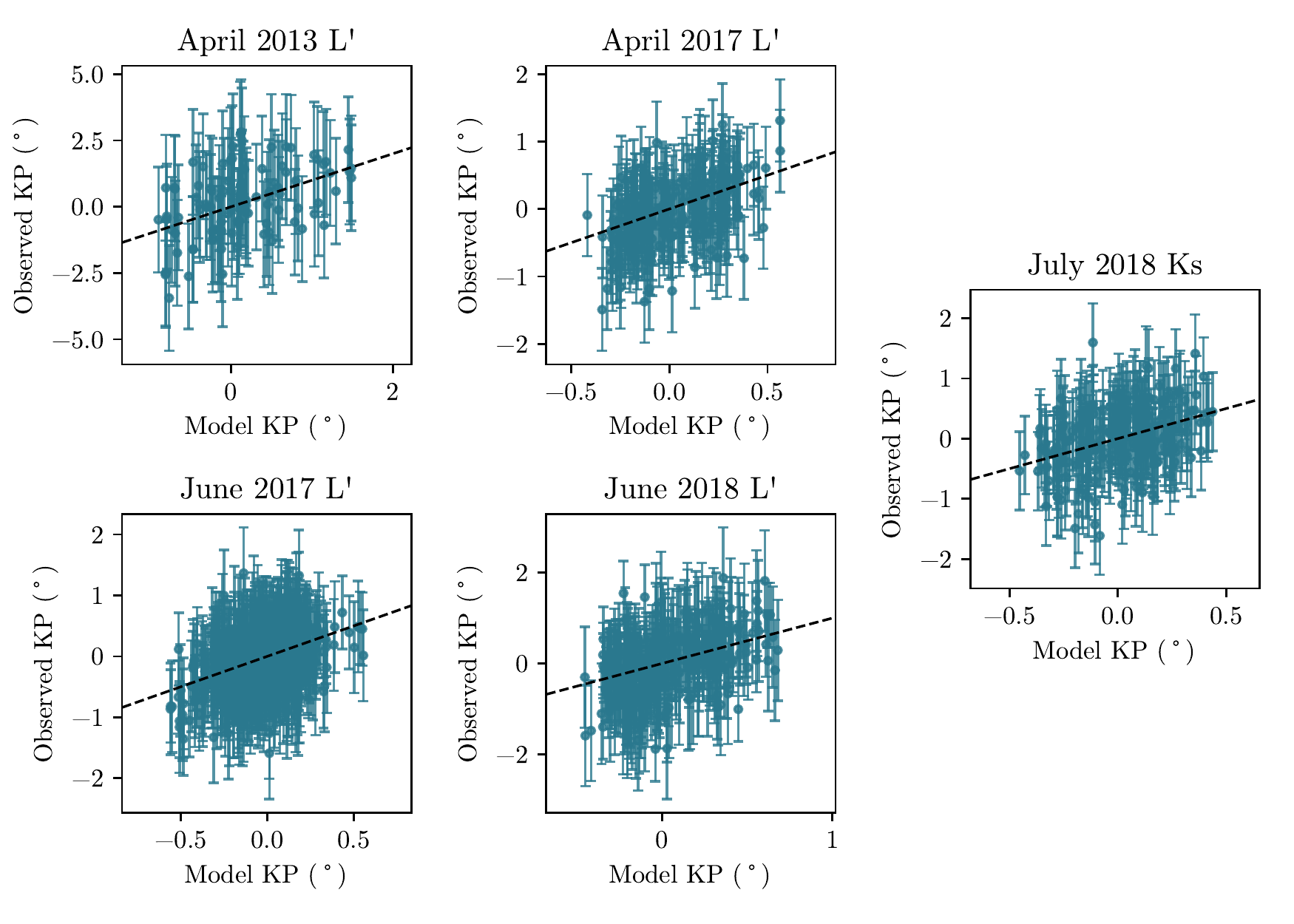}
\end{tabular}
\end{center}
\caption
{ \label{fig:kpcomp}
Observed kernel phases with error bars, versus single companion model kernel phases for each observational epoch. The dashed line indicates a 1:1 scaling.
}
\end{figure*}

\begin{figure*}[ht]
\begin{center}
\begin{tabular}{c} \includegraphics[width=0.8\textwidth]{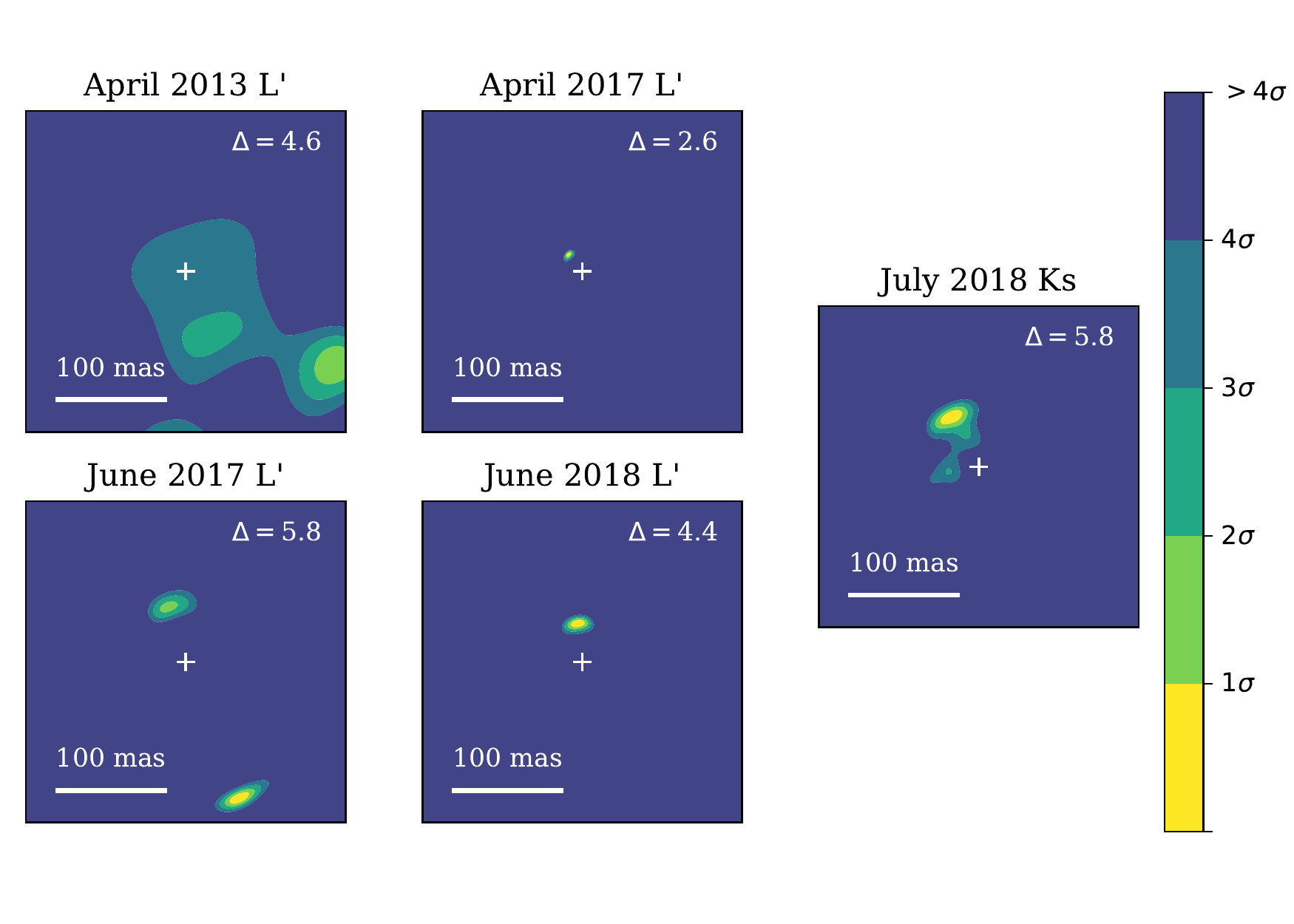}
\end{tabular}
\end{center}
\caption
{ \label{fig:binorb}
Individual panels show $\chi^2$ slices at the best fit contrast ratio for MagAO NRM observations (top left), and Keck NRM observations (all other panels). The filled contours show $1-5\sigma$ significance levels (e.g. yellow regions are within 1$\sigma$ of the best fit companion model).
}
\end{figure*}

We fit single companion models to the kernel phases using a grid search.
The models consist of a central delta function and a second delta function having a separation $s$, position angle PA, (measured east of north), and contrast $\Delta$. 
We sample the grid in single degree intervals for PA, every 2 mas in $s$ (out to 750 mas), and every 0.1 magnitudes (up to 10 mag) in $\Delta$.
Since any orbital motion at resolved angular separations ($\sim23$ mas at K$\mathrm{_s}$ and $\sim40$ mas at L$'$) would be small on the timescale of one night, we fit the June 6 -- 7, 2017 data together and the June 28 -- 29, 2018 data together.
Table \ref{tab:comps} lists the best fit model for each epoch, Figure \ref{fig:kpcomp} shows the observed versus best fit model kernel phases, and Figure \ref{fig:binorb} shows $\chi^2$ slices at the best fit companion contrast for each epoch.

\begin{deluxetable*}{lccccccc}
\caption{Single Companion Fit Results\label{tab:comps}}
\tablehead{\colhead{Dataset} & \colhead{PA (deg)} & \colhead{s (mas)} & \colhead{$\Delta$ (mag)} & \colhead{$\Delta \chi^2$} & \colhead{Sig.\tablenotemark{a}} & \colhead{FPP$1$\tablenotemark{b}}& \colhead{FPP$2$\tablenotemark{c}}}
\startdata
April 2013 L$'$ & $314\pm2$ & $488\pm^{12}_{16}$ & $4.6\pm^{0.4}_{0.2}$ & 25.8 & 5$\sigma$& 0.89 & 0.005\\
April 2017 L$'$ & $40\pm6$ & $20\pm^{28}_{4}$ & $ 2.6 \pm^{2.6}_{2.4}$& 70.2 &$>5\sigma$ & $<0.05$ & $<0.00002$ \\
June 2017 L$'$ & $202\pm^{178}_{180}$ & $136\pm^4_{12}$ & $5.8\pm^{0.2}_{5.6}$ & 107.2 & $>5\sigma$& $<0.008$ & $<0.00004$  \\
June 2018 L$'$ & $6\pm^{10}_8$ & $36\pm^{20}_{16}$ & $4.4\pm^{1.2}_{3.8}$ & 71.2 & $>5\sigma$& $<0.067$ & $<0.00005$ \\
July 2018 $\mathrm{K_s}$ & $28\pm{12}_{10}$ & $52\pm4$ & $5.8\pm^{0.4}_{0.2}$ & 33.8 & $>5\sigma$& 0.098 & 0.0004\\
\enddata
\tablenotetext{a}{Significance with which the single companion model is preferred over the null (single point source) model, using the $\Delta \chi^2$ values listed in this table.}
\tablenotetext{b}{False positive probability derived from distribution of best fit contrasts to Gaussian noise.}
\tablenotetext{c}{False positive probability derived from distribution of best fit F-statistics for Gaussian noise.}
\end{deluxetable*}

\begin{figure*}[ht]
\begin{center}
\begin{tabular}{c} \includegraphics[width=\textwidth]{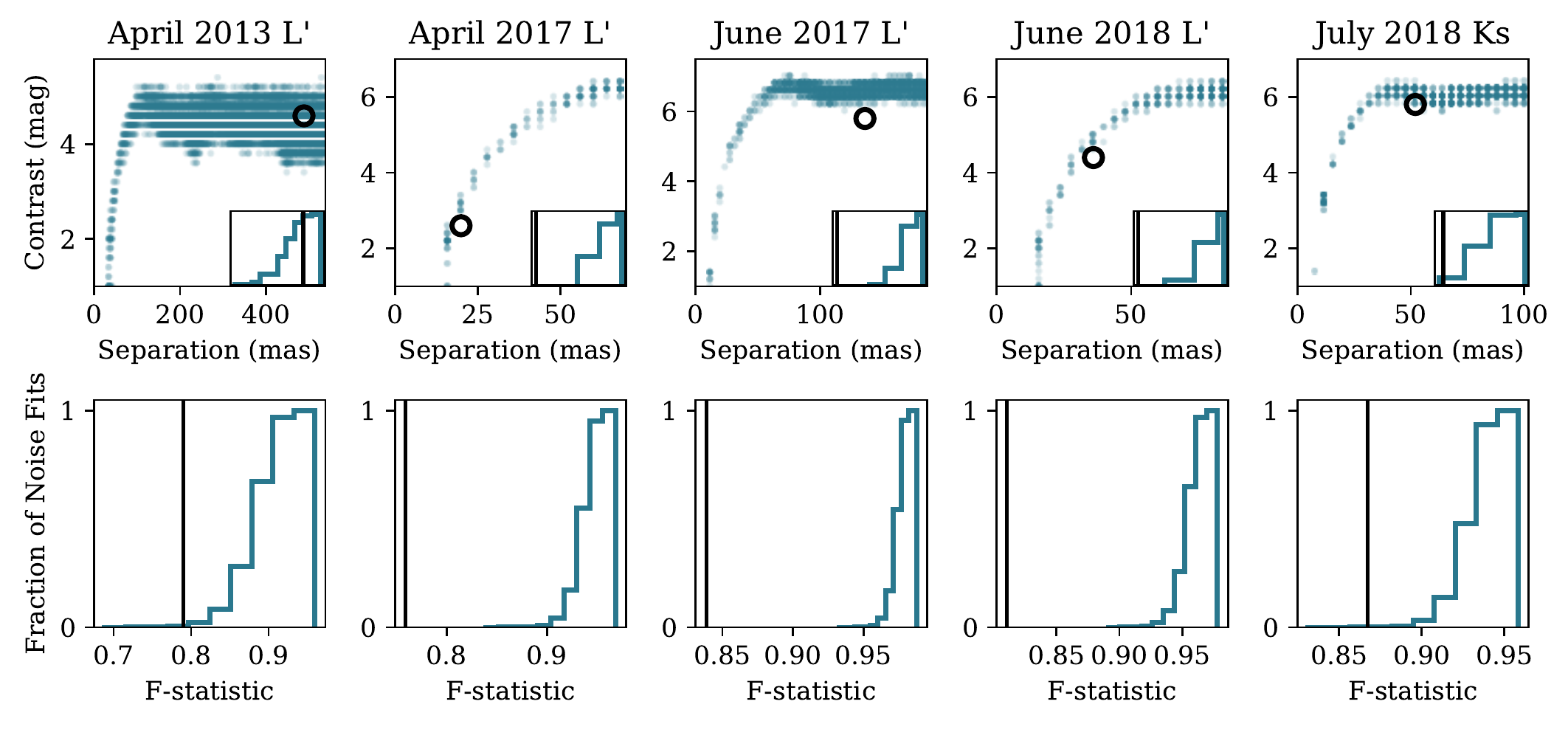}
\end{tabular}
\end{center}
\caption
{ \label{fig:fafig}
Top: Scattered points in the large panels show single companion fits to Gaussian noise realizations drawn from the distributions shown in Figure \ref{fig:hists}. The hollow circles show the best fits to the data for each epoch. Each inset panel shows the cumulative histogram of best fit contrasts for noise fits with the same separation as the fit to the data. The vertical line shows the best fit contrast for the data. We calculate the false positive probability for each epoch by finding the fraction of noise fits with equal or lower contrast than the best fit contrast for the data.
Bottom: Histograms show the distributions of F-statistics for fits to Gaussian noise realizations drawn from the distributions shown in Figure \ref{fig:hists}. The vertical lines show the best fit F-statistic for the single companion models listed in Table \ref{tab:comps}. For each epoch, the fraction of fits to noise that have F-statistics lower than the best fit companion yields a false positive probability.
}
\end{figure*}

We calculate the $\chi^2$ interval between the best fit companion model and the null model to estimate the significance of the best fit relative to the null model.
Since $\chi^2$ intervals can be corrupted by poor error estimation, we also use Monte Carlo methods to estimate the detection significance. 
We fit a single companion model to many ($> 10^{5}$) Gaussian noise realizations drawn from the distributions shown in Figure \ref{fig:hists}.
We use these fits to calculate the false positive probability for each epoch in two ways: 
(1) We find the number of noise fits that have the same separation as the best fit to the data, and take the false positive probability to be the number of noise fits with equal or lower contrast than the fit to the data \citep[e.g.][]{2011ApJ...731....8K}.
(2) We find the distribution of F-statistics, the best fit $\chi^2$ divided by the null model $\chi^2$, and calculate the fraction of F-statistics for noise that are less than the F-statistic for the data \citep[e.g.][]{2002ApJ...571..545P,2015ApJ...801...85S}. 
Figure \ref{fig:fafig} shows the distributions of fits to simulated kernel phase noise as well as the distributions of F-statistics. 
Table \ref{tab:comps} lists the $\chi^2$ intervals, the corresponding significance levels, and the false positive probabilities.

\begin{figure}[h!]
\begin{center}
\begin{tabular}{c} 
\includegraphics[width=3in]{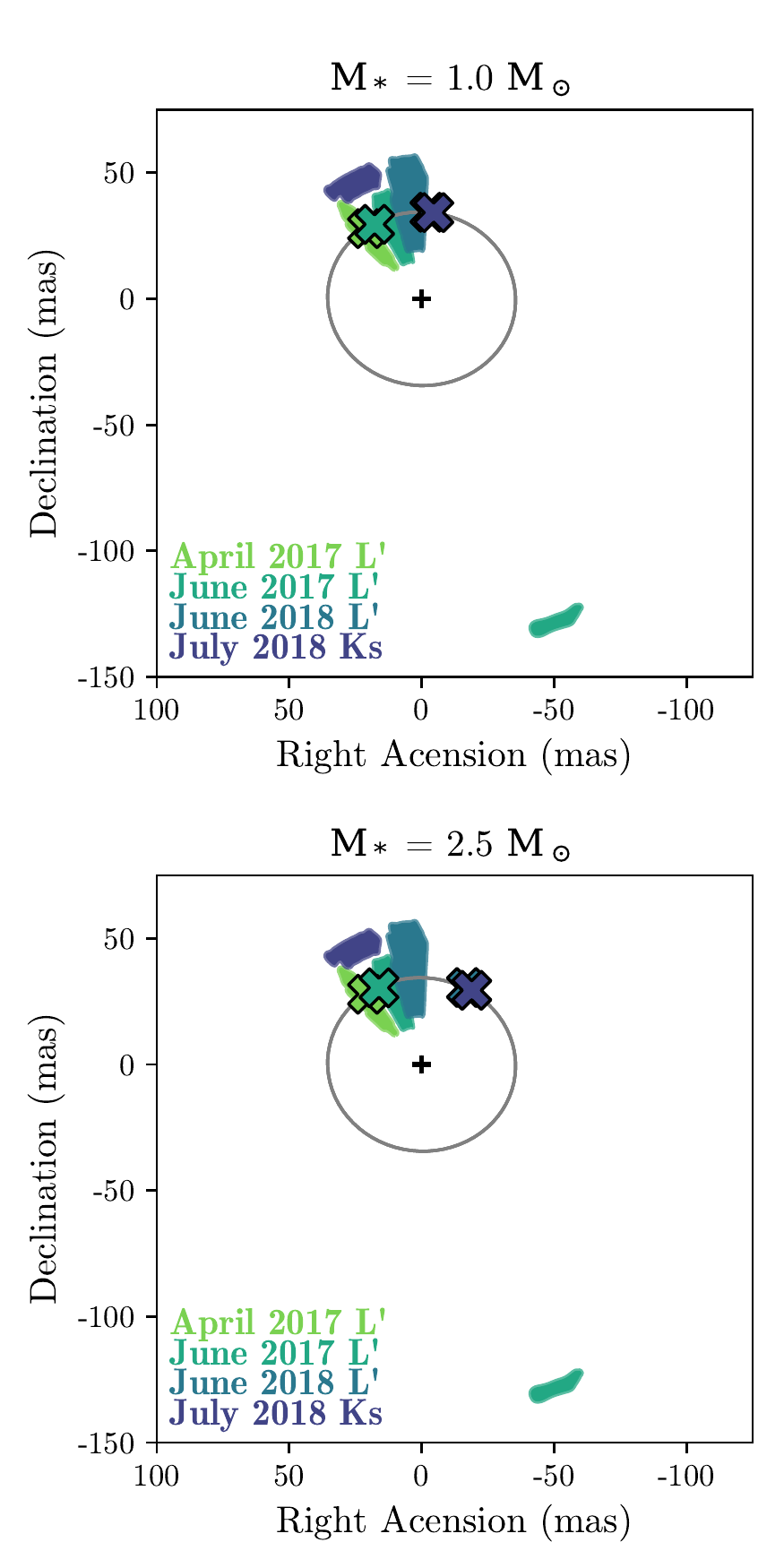}
\end{tabular}
\end{center}
\caption
{ \label{fig:comporb} 
Shaded regions show 1$\sigma$ allowed positions for the four NRM datasets with kernel phase companion fits that are inconsistent with noise. The scattered x's show predicted orbital positions during each epoch, for Keplerian orbits aligned with the millimeter disk starting roughly at the position of the April 2017 best companion fit. The grey line traces out the entirety of the orbit. The top panel shows predictions for a stellar mass of 1 $\mathrm{M_\odot}$, and the bottom panel for 2.5 $\mathrm{M_\odot}$. The multi-epoch observations are not consistent with orbital motion in the outer disk plane.
}
\end{figure}

Since statistically-significant companion model fits exist, we investigate whether they can be explained by an orbiting companion in the plane of the outer disk (Figure \ref{fig:comporb}), for both of SR 21's stellar mass estimates ($\mathrm{1~M_\odot}$ and $\mathrm{2.5~M_\odot}$).
The best fit single-companion models do not have positions consistent with orbital motion in the disk plane.
The June 2017 best fit, which lies southwest of the star, is $\sim180^\circ$ offset from the April 2017 and June 2018 best fits.
While companions to the northeast of the star are allowed by all datasets at 1$\sigma$, the $1\sigma$ allowed regions are inconsistent with orbital motion in the disk plane for both stellar mass estimates (Figure \ref{fig:comporb}).
Extended, static brightness distributions (often associated with scattered light from disk material) have been known to cause spurious companion phase signals in single-epoch datasets \citep[e.g.][]{2011A&A...528L...7H,2015ApJ...801...85S,2015MNRAS.450L...1C}, and position angles that change erratically between local $\chi^2$ minima in multi-epoch datasets \citep[e.g.][]{2016SPIE.9907E..0DS}.
The observed asymmetries in the multi-epoch dataset presented here are more consistent with a static, extended structure than with an orbiting companion.

\begin{figure*}[ht]
\begin{center}
\begin{tabular}{c} 
\includegraphics[width=0.8\textwidth]{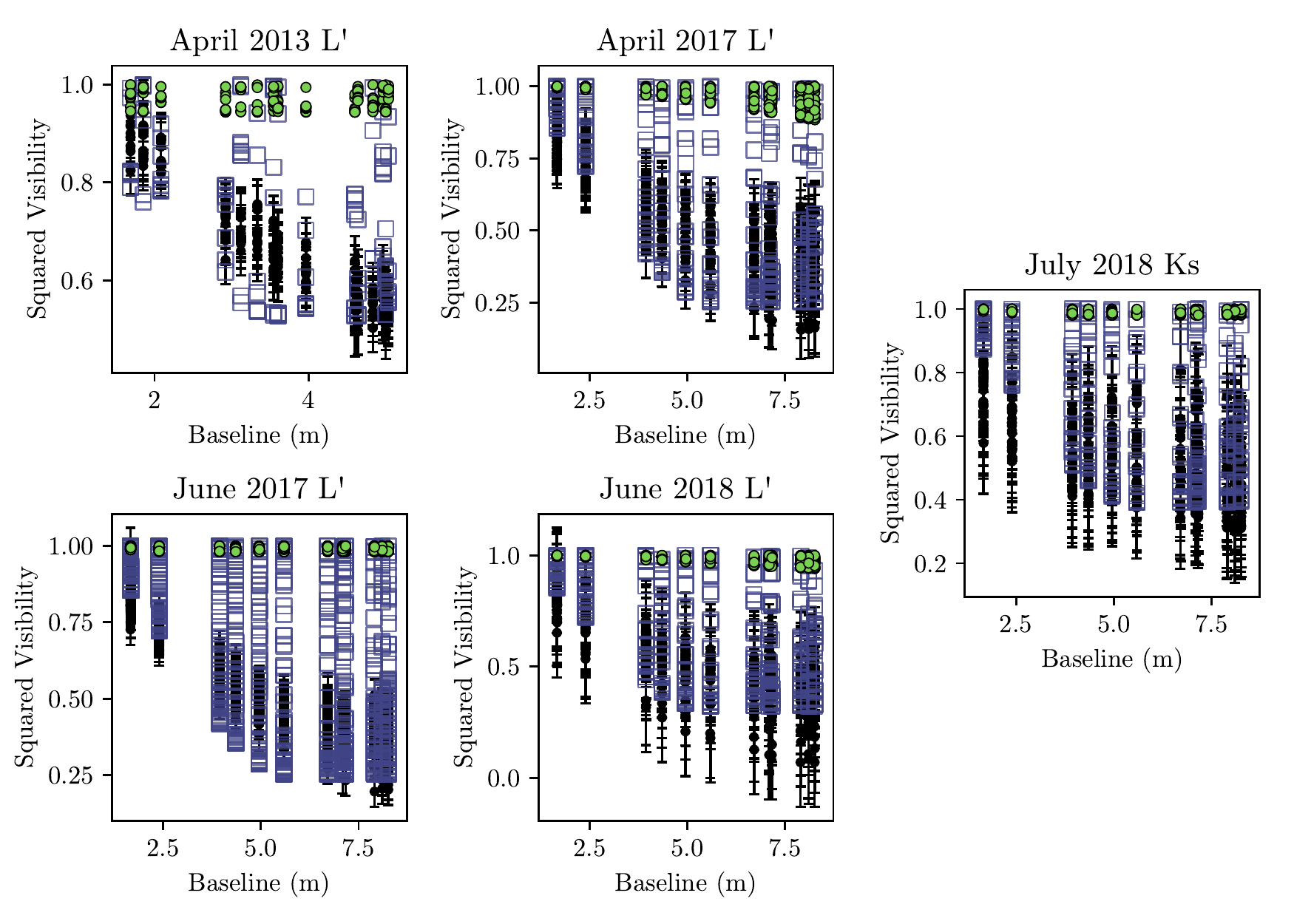}
\end{tabular}
\end{center}
\caption
{ \label{fig:v2compkp}
Black points with error bars show the observed squared visibilities for each epoch. Green scattered circles show the model squared visibilities for the single companion fit to the kernel phases. Blue scattered squares show the model squared visibilities for a single companion fit to the squared visibilities alone.
}
\end{figure*}

The squared visibilities are also inconsistent with a simple single-companion model.
The companion models that reproduce the kernel phases over-predict the squared visibilities; the models are under-resolved compared to the observations. 
Furthermore, the squared visibilities have consistent values over the entire range of parallactic angles in each dataset, suggesting a more symmetric morphology than a single companion model.
Indeed, single companion fits to the squared visibilities alone cannot reproduce them.
Figure \ref{fig:v2compkp} shows this: to match the lowest visibilities, a binary model would require low contrast.
In order to match the uniformity with parallactic angle, however, the brightness distribution must be relatively centro-symmetric. 
Both the kernel phases and the squared visibilities point toward a relatively static, extended brightness distribution.

\subsection{Geometric Disk Models}
Since we do not observe orbital motion in the single companion fits, and since a simple companion model cannot reproduce the visibilities, we next fit a grid of geometric disk models to the data to estimate the size of the near infrared emission.
Assuming a static disk, we fit all epochs at each wavelength simultaneously.
Each model consists of a 2-dimensional Gaussian with an axes ratio ($r$) and major axis position angle ($\theta$; measured east of north), as well as a central $\delta$ function accounting for some fraction $f$ of the total image flux \citep[e.g.][]{2017ApJ...844...22S}.
We use a $\chi^2$ interval to derive the parameter uncertainties.
Due to their centro-symmetry, these models all predict zero kernel and closure phases.
We first fix the axes ratio at 1, ignoring the major axis position angle, and we then let $r$ and $\theta$ vary; Table \ref{tab:gaussfits} lists the results.
The circular Gaussian models prefer similar sizes for the L$'$ and $\mathrm{K_s}$ emission, while the elliptical Gaussian models prefer a smaller axis ratio and larger FWHM for $\mathrm{Ks}$ compared to L$'$, with nearly-overlapping parameter error bars.
In both cases, the central delta function flux is significantly higher at $\mathrm{Ks}$ than at L$'$.

\begin{deluxetable}{lcc}
\caption{Geometric Disk Fit Results\label{tab:gaussfits}}
\tablehead{\colhead{Parameter} & \colhead{Keck L$'$} & \colhead{Keck $\mathrm{K_s}$}}
\startdata
\multicolumn{3}{c}{\textbf{Circular Gaussian}}\\
$FWHM$ (mas) & 110.7 $\pm^{0.3}_{1.4}$ & 111 $\pm^{10}_{7}$\\
$f$ & 0.647 $\pm$ 0.002 & 0.715 $\pm^{0.005}_{0.007}$\\
\multicolumn{3}{c}{\textbf{Elliptical Gaussian}}\\
$FWHM$ (mas) & 113.6$\pm^{0.4}_{3}$ & 130 $\pm^{10}_{20}$\\
$\theta$ ($^\circ$) & 160 $\pm$ 10& 90 $\pm^{30}_{20}$\\
$r$ & 0.95 $\pm^{0.03}_{0.02}$  & 0.7 $\pm^{0.2}_{0.1}$\\
$f$ & 0.647 $\pm$ 0.002 & 0.714 $\pm^{0.006}_{0.009}$ \\
\enddata
\end{deluxetable}

\subsection{Radiative Transfer Disk Models}
\subsubsection{Aligned Models}\label{sec:aligned}
SR 21 is known to have a small-grain disk that extends within the millimeter clearing \citep[e.g.][]{2009ApJ...698L.169E,2013ApJ...767...10F}.
To investigate whether a small-grain circumstellar disk aligned with the millimeter disk can reproduce the observations, we fit a grid of radiative transfer models to the imaging data and to a spectral energy distribution gathered from the literature.
We include the same photometry as \citet{2013ApJ...767...10F}, from the Spitzer Infrared Array Camera (IRAC) and Multiband Imaging Photometer (MIPS), AKARI, Infrared Astronomical Satellite (IRAS), the Submillimeter Array (SMA), and the James Clerk Maxwell Telescope Submillimetre Common-User Bolometer Array (SCUBA).
We also add Herschel PACS and SPIRE photometry from 70 $\mu$m to 500 $\mu$m \citep{2017ApJ...849...63R}.

We use the radiative transfer software, \texttt{RADMC-3D} \citep{2012ascl.soft02015D} to generate synthetic disk images and spectra self-consistently.
We use the \texttt{python} library, \texttt{pdspy} \citep{patrick_sheehan_2018_2455079,2019ApJ...874..136S} to generate the necessary input files for \texttt{RADMC-3D} (e.g. dust density grids and radiation sources), to call \texttt{RADMC-3D} to do thermal calculations and generate a large grid of model images and spectra, and to read the \texttt{RADMC-3D} output files.
For each disk, we input a standard density profile for a flared disk:
\begin{equation}
\rho\left(r,z\right) = \rho_0 \left(\frac{r}{r_0}\right)^{-\alpha} \exp\left(-\frac{1}{2}\left[\frac{z^2}{h\left(r\right)}\right]^2\right),
\end{equation}
where
\begin{equation}
h\left(r\right) = h_0 \left(\frac{r}{r_0}\right)^\beta
\end{equation}
gives the disk scale height. 
The values $r$ and $z$ are the radius and height in cylindrical coordinates, $h_0$ and $\rho_0$ represent the scale height and density at some reference radius $r_0$ (taken to be 1 AU), and $\alpha$ and $\beta$ are the density and scale height power law indices. 
The disk surface density is given by:
\begin{equation}\label{eq:sig}
\Sigma = \Sigma_0 \left(\frac{r}{r_0}\right)^{-\gamma},
\end{equation}
where $\gamma = \alpha - \beta$, and $\Sigma_0$ is the surface density at $r_0$.
$\Sigma_0$ and $\rho_0$ can be found by assuming a total disk mass and integrating the disk density over all space.

For each model we include two disk components: (1) a relatively thin large-grain disk, and (2) a puffier small-grain disk.
We fix the total disk dust mass to $6\times10^{-5}~\mathrm{M_\odot}$  \citep[e.g.][]{2011ApJ...732...42A}.
Each disk component has the following parameters: fractional mass $f_m$, inner radius $r_i$, outer radius $r_o$, minimum grain size $a_{min}$, maximum grain size $a_{max}$, grain size distribution index $p$, as well as $\gamma$, $\beta$, and $h_0$.
For the large-grain disk, we fix $f_m$, $r_{i,l}$, $r_{o,l}$, $\gamma_l$, $\beta_l$, and $h_{0,l}$ to values consistent with the literature \citep{2011ApJ...732...42A,2009ApJ...704..496B,2014ApJ...783L..13P,2016A&A...585A..58V}.
We note that ALMA observations at 450 $\mu$m and 870 $\mu$m lead to different derived cavity radii \citep[34 and 41 AU, respectively][]{2015A&A...584A..16P}, but the two component disk model cannot reproduce this difference. 
We thus assume a single intermediate value (36 AU) for $r_{i,l}$, and checked that changing it between 34 and 41 AU did not change the results significantly.

We allow all the disk geometry parameters for the small grain disk to vary, and we also vary the minimum and maximum grain sizes for both disk components, requiring that the minimum grain size for the large grain disk is not smaller than the minimum grain size in the small grain disk (see Table \ref{tab:diskgrids}).
We include non-hydrostatic values for the small-grain disk flaring index, $\beta_s$, to allow for the possibility of a puffed-up inner rim and/or a disk wind without adding additional model components.
We fix the disk inclination to $i = 15^\circ$, and test two position angle values consistent with CO observations \citep[$PA$ = 14$^\circ$ and 194$^\circ$;][]{2008ApJ...684.1323P}.\footnote{Changing $PA$ by 180$^\circ$ simply reverses the near- and far-side of the disk, which were not constrained by previous studies.}
Following \citet{2011ApJ...732...42A}, we fix the fractional mass of the small-grain disk to $0.15$, and for both disk components we fix the dust grain size distribution index, $p$, to 3.5. 
We adopt a similar dust composition as previous studies - 65\% silicates and 35\% graphite \citep[e.g.][]{2001ApJ...548..296W}.

\begin{deluxetable}{lc}
\tablecaption{Aligned Disk Model Grid\label{tab:diskgrids}}
\tablehead{\colhead{Parameter} & \colhead{Values Explored}}
\startdata
$T_{*}$ (K) & 5750\tablenotemark{*} \\
$R_{*}$ (R$_\odot$) & 3.15$^*$ \\
$i$ ($^\circ$) & 15$^{*}$\\
$PA$ ($^\circ$) & 14$^{*}$, 194$^{*}$\\
$p$ & 3.5$^*$\\
\hline
\multicolumn{2}{l}{\textbf{Large Grain Disk}}\\
$f_{m,l}$ & 0.85\tablenotemark{*}\\
$r_{i,l}$ (AU) & 36\tablenotemark{$\ddagger$}\\
$r_{o,l}$ (AU) & 200$^*$\\
$\gamma_l$ & 1.0$^*$\\
$\beta_l$ & 1.15$^*$\\
$h_{0,l}$\tablenotemark{$\dagger$} (AU) & 0.0076$^*$\\
$a_{min,l}$ ($\mu$m) & 0.1, 1.0, 10.0, 100.0\\
$a_{max,l}$ ($\mu$m) & 1000.0$^*$, 10000.0\\
\hline
\multicolumn{2}{l}{\textbf{Small Grain Disk}}\\
$f_{m,s}$ & 0.15\tablenotemark{*}\\
$r_{i,s}$ (AU) & 0.07, 3.0, 4.0, 5.0, 6.0, 7.0\\
$r_{o,s}$ (AU) & 200$^*$\\
$\gamma_s$ & 0.0, 0.5, 1.0$^*$\\
$\beta_s$ & 1.15$^*$, 1.4, 1.65, 1.9, 2.15, 2.4, 2.65\\
$h_{0,s}$\tablenotemark{$\dagger$} (AU) & 0.025, 0.05, 0.075, 0.1, 0.15, 0.2, 0.25, 0.3\\
$a_{min,s}$ ($\mu$m) & 0.1, 1.0, 2.0, 5.0\\
$a_{max,s}$ ($\mu$m) & 10.0$^*$\\
\enddata
\tablenotetext{*}{Values consistent with the literature.}
\tablenotetext{\dagger}{We define $h_0$ as the disk scale height at 1 AU.}
\tablenotetext{\ddagger}{The assumed large grain inner radius of 36 AU is intermediate between the ALMA estimate at 450 $\mu$m (34 AU) and the SMA and ALMA estimates at 870 -- 880 $\mu$m (40-41 AU).}
\end{deluxetable}

Since the imaging and SED are different datasets with different error bars, following \citet{2017ApJ...851...45S}, we create a goodness-of-fit metric ($X^2$) that combines the kernel phase, squared visibility, and SED $\chi^2$ values with relative weights:
\begin{equation}
X^2 = w_{KP} \frac{\chi^2_{KP}}{\chi^2_{min,KP}} +  w_{V^2} \frac{\chi^2_{V^2}}{\chi^2_{min,V^2}} + w_{SED} \frac{\chi^2_{SED}}{\chi^2_{min,SED}}.
\end{equation}
We normalize the $\chi^2$ arrays by their minimum values, to ensure that we can easily explore weights that both up- and down-weight each dataset relative to the others.
This arbitrary choice does not affect the final best fit, since the same relative weighting could be found without normalizing any of the $\chi^2$ values by a scalar.
We explore a wide range of weights ($\sim 0.005 - 10$) for each dataset.

\begin{figure*}[ht]
\begin{center}
\begin{tabular}{c}
\includegraphics[width=\textwidth]{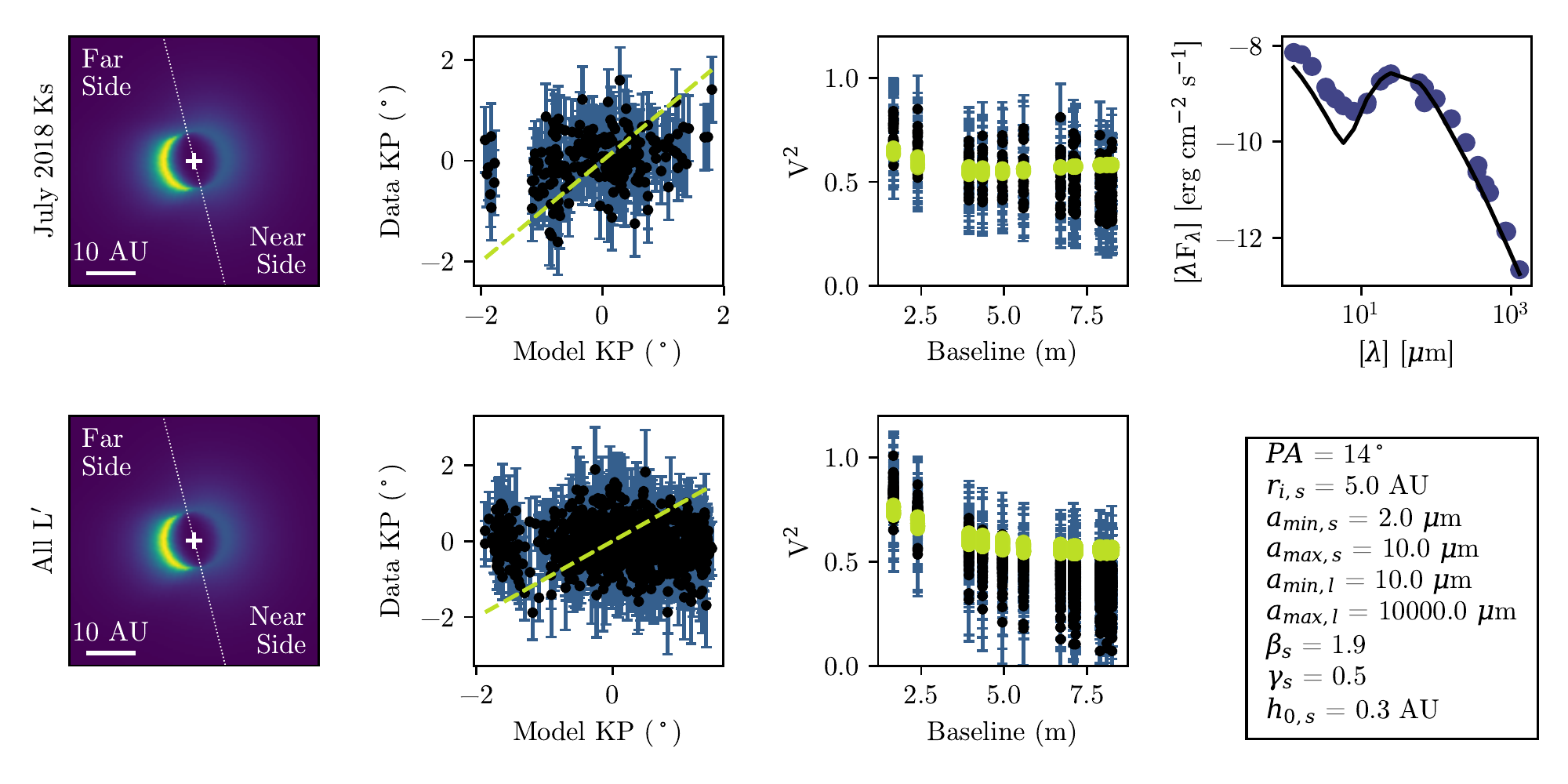}
\end{tabular}
\end{center}
\caption
{ \label{fig:disk1}
Left: Example radiative transfer images for models with a small-grain disk that is aligned with the millimeter disk. The top and bottom panels show $\mathrm{K_s}$ and L$'$, respectively. White crosses mark the position of the star, and the white dotted line shows the orientation of the disk major axis. Labels indicate the near and far sides of the disk, given the inclination of 15$^\circ$. Center left: observed versus disk model kernel phases for the 2018 July $\mathrm{K_s}$ dataset (top) and all L$'$ observations (bottom). The yellow dashed line indicates a 1:1 relation. Center right: Black points with error bars show the squared visibilities for the 2018 July $\mathrm{K_s}$ dataset (top) and all L$'$ observations (bottom). The yellow points show the disk model visibilities. Right: Blue points show the spectral energy distribution gathered from the literature, and the black line shows the disk model SED.}
\end{figure*}

Figure \ref{fig:disk1} shows the best fit disk model and its parameters for $w_{SED}$ = 0.5,  $w_{KP}$ = 5, and  $w_{V^2}$ = 1. 
The aligned disk model can roughly reproduce the SED and the squared visibilities, but cannot match the kernel phases.
The second panels from the left in Figure \ref{fig:disk1} show this; there is a slope offset between the 1:1 model-vs-data relation and the plotted points, especially at L$'$.
All asymmetries in this set of models are along the wrong position angle compared to the data.

The models that most closely match the observations have the following characteristics: (1) an inner disk truncation at $\sim4-7$ AU, (2) a high flaring index ($\beta \sim 2$), and thus a large disk scale height at the truncation radius, and (3) minimum grain sizes larger than $\sim 2-5~\mu$m.
The disk models with inner radii at the sublimation radius over-predict the squared visibilities; the hot dust close to the star creates too much unresolved flux.
Models with many small grains and/or optically-thick disk rims can match the imaging data, but over-predict the 10 $\mu$m silicate feature in the SED.
Disks with large scale heights and flaring indices scatter enough light to match the angular size of the $\mathrm{K_s}$ and L$'$ imaging, as well as the magnitude (but not direction) of the asymmetry, without over-predicting the 10$\mu$m SED.
However, as shown in Figure \ref{fig:disk1}, the large scale height can lead to an under-prediction of the short wavelength spectral energy distribution. 
These tests suggest that more complex disk models are required to reproduce the data.

\subsubsection{Warped Disk Models}
Since the aligned disk models cannot match the imaging, here we explore a set of warped disk models.
We allow the inner regions of the small-grain disk to have both a different inclination and position angle from the millimeter disk.
We follow a procedure similar to that presented in \citet{2018MNRAS.477.5104C}, where the small grain disk has an initial inclination and position angle at its inner radius, a linearly changing orientation through intermediate radii, and the same orientation as the millimeter disk at larger radii.
We force the small grain disk to have the same orientation as the millimeter disk by a stellocentric radius of 7 AU, in order to be consistent with the gas position angle determined through spectro-astrometry \citep{2008ApJ...684.1323P}.

We first fit a coarse grid of warped disk models to roughly estimate the best initial inclination and position angle. 
We fix the small-grain inner radius and density structure to be the same as that shown in Figure \ref{fig:disk1}.
We vary the initial position angle and inclination from 0$^\circ$ to 350$^\circ$ and $0^\circ$ to $80^\circ$, respectively, both with grid spacings of $10^\circ$.
We force the disk orientation to change linearly to $i = 15^\circ$ and either $PA = 14^\circ$ or $PA = 194^\circ$ between 5 AU and 7 AU.
The best fit orientation has an initial position angle of $280^\circ$ and initial inclination of $20^\circ$, and a final position angle of $PA = 194^\circ$.

We next fit a larger grid of warped disk models informed by the initial fits (see Table \ref{tab:warpgrid}).
We vary the initial position angle and inclination around the best fit values from the coarse grid, and set the disk orientation to change linearly starting at a fraction $f_t$ of the disk inner radius ($r_{i,s}$).
At a disk radius of $r_1 = 7$ AU, the final position angle and inclination are $PA_1 = 194^\circ$ and $i_1 = 15^\circ$.
We once again vary $\beta$, $\gamma$, and $h_0$, and now include a range of values for the small-grain disk mass.
We introduced this free parameter to allow for different optical depths for disks with the same geometries and dust properties.

\begin{deluxetable}{lc}
\tablecaption{Warped Disk Model Grid\label{tab:warpgrid}}
\tablehead{\colhead{Parameter} & \colhead{Values Explored}}
\startdata
$T_{*}$ (K) & 5750\tablenotemark{*} \\
$R_{*}$ (R$_\odot$) & 3.15$^*$ \\
$p$ & 3.5$^*$\\
\hline
\multicolumn{2}{l}{\textbf{Large Grain Disk}}\\
$\mathrm{M_{d,l}}~(\mathrm{M_\odot}$) & 5.1e-5\\
$r_{i,l}$ (AU) & 36\tablenotemark{$\ddagger$}\\
$r_{o,l}$ (AU) & 200$^*$\\
$\gamma_l$ & 1.0$^*$\\
$\beta_l$ & 1.15$^*$\\
$h_{0,l}$\tablenotemark{$\dagger$} (AU) & 0.0076$^*$\\
$a_{min,l}$ ($\mu$m) & 10.0\\
$a_{max,l}$ ($\mu$m) &10000.0\\
\hline
\multicolumn{2}{l}{\textbf{Warped Small Grain Disk}}\\
$\mathrm{M_{d,s}}~(\mathrm{M_\odot}$) & $[0.025,0.03,0.04,0.05, 0.1, 0.15^*, 0.2] \times$ 6.0e-5\\
$r_{i,s}$ (AU) & 3.0, 4.0, 4.1, 4.2, 4.3, 4.4, 4.5, 4.6, 4.7, 4.8, 4.9\\
& 5.0, 5.1, 5.2, 5.3, 5.4, 5.5, 5.6, 5.7, 5.8, 5.9, 6.0\\
$r_{o,s}$ (AU) & 200$^*$\\
$PA_{0}$ ($^\circ$) & 270, 280, 290, 300, 310, 320, 330, 340\\
$i_{0}$ ($^\circ$) & 15, 25, 35, 40\\
$PA_{1}$ ($^\circ$) & 194\\
$i_{1}$ ($^\circ$) & 15\\
$f_{t}$ & 1.0, 1.05, 1.1, 1.15, 1.2, 1.25, 1.3, 1.35, 1.4, 1.45, 1.5\tablenotemark{$\P$} \\
$r_{1}$ (AU) & 7.0 \\
$\gamma_s$ & 0.5, 0.7\\
$\beta_s$ & 1.15, 1.35, 1.55, 1.75, 1.95\\
$h_{0,rin}$\tablenotemark{$\dagger$} (AU) & 0.3, 0.4, 0.5, 0.75, 1.0, 2.0, 2.5, 3.0, 3.5, 4.0, 5.0\\
$a_{min,s}$ ($\mu$m) & 2.0, 5.0\\
$a_{max,s}$ ($\mu$m) & 10.0$^*$\\
\enddata
\tablenotetext{*}{Values consistent with the literature.}
\tablenotetext{\dagger}{Since the disk scale height at the inner radius has a large effect on the kernel phase signal, we re-parameterized this set of models in terms of $h_0$ at the inner disk radius.}
\tablenotetext{\ddagger}{The assumed large grain inner radius of 36 AU is intermediate between the ALMA estimate at 450 $\mu$m (34 AU) and the SMA and ALMA estimates at 870 -- 880 $\mu$m (40-41 AU).}
\tablenotetext{\P}{We only explore grid points where $f_t \times r_{i,s}$ is less than 7 AU.}

\end{deluxetable}

\begin{figure*}[ht]
\begin{center}
\begin{tabular}{c}
\includegraphics[width=\textwidth]{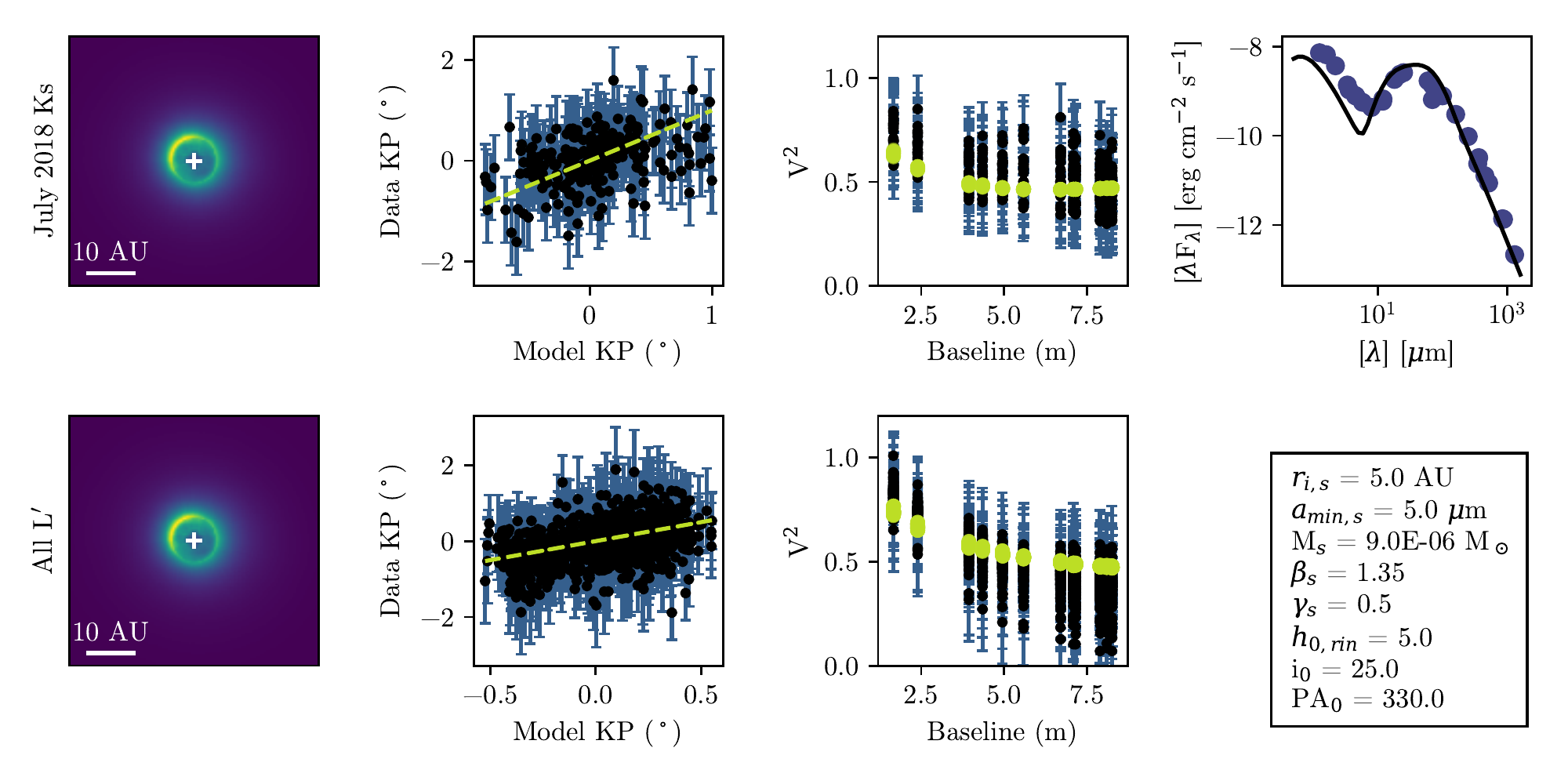}
\end{tabular}
\end{center}
\caption
{ \label{fig:diskw}
Left: Example radiative transfer images for models with a small-grain disk that has a changing orientation relative to the millimeter disk. The top and bottom panels show $\mathrm{K_s}$ and L$'$, respectively. Center left: observed versus disk model kernel phases for the 2018 July $\mathrm{K_s}$ dataset (top) and all L$'$ observations (bottom). The yellow dashed line indicates a 1:1 relation. Center right: Black points with error bars show the squared visibilities for the 2018 July $\mathrm{K_s}$ dataset (top) and all L$'$ observations (bottom). The yellow points show the disk model visibilities. Right: Blue points show the spectral energy distribution gathered from the literature, and the black line shows the disk model SED.}
\end{figure*}

We select the best fit warped disk models using the same weighting scheme as the aligned disk models.
Figure \ref{fig:diskw} shows representative images, squared visibilities and kernel phases, and the spectral energy distribution.
The warped disk models can better match the observables, in particular the kernel phases.
Like the aligned models, the warped models prefer large grains to reproduce both the imaging data and the small 10 $\mu$m feature in the spectral energy distribution.
They prefer geometries with large scale heights, even when the small grain disk mass is allowed to vary.
To check whether the warped disks also prefer an inner rim at a radius of a few AU, we ran a small grid of warped models with inner radii of 0.07 AU.
Like the aligned models, warped models that extend to the sublimation radius over-predict the squared visibilities.

We also checked that the warped disk model is consistent with the MagAO H$\alpha$  + continuum images, by KLIP processing a datacube of the model convolved with a PSF having FWHM = 17 pixels. 
The signal in the final KLIP-processed image is lower than the noise in the observed H$\alpha$  + continuum image. 
Directly injecting the PSF-convolved disk model into the H$\alpha$  + continuum data did not significantly alter the KLIP-processed image.

\subsection{Parametric Imaging}
To look for complex structure beyond simple disk models, we reconstruct images with SQUEEZE \citep{2010SPIE.7734E..2IB}, which uses Markov-Chain Monte Carlo methods to sample the image posterior distribution.
We run SQUEEZE in parallel-tempering mode, where chains with different acceptance probabilities can exchange information, in order to effectively explore the parameter space. 
SQUEEZE can include several regularizers and model components; we explore a variety of these options for the image reconstructions.

SQUEEZE can reconstruct images while simultaneously fitting parametric model components such as a central delta function or a resolved disk.
We reconstruct images including both of these components - a central delta function representing the star, and a central resolved, circularly-symmetric disk to account for the drop in the observed squared visibilities. 
These components have three free parameters: (1) the fractional image flux in the delta function, (2) the outer radius of the circularly-symmetric disk, and (3) the fractional image flux in the circularly-symmetric disk.
SQUEEZE is free to put 0\% of the flux in the resolved disk component; including this component does not force the image to be symmetric, but it can constrain the overall size of the structure in the reconstructed image.

We tested three separate regularizations: (1) maximum entropy \citep{1972JOSA...62..511F}, which favors images with minimal configurational information; (2) total variation \citep{1992PhyD...60..259R}, which prefers images where most gradients are zero (favoring smooth, piecewise images); and (3) the $l-0$ norm, \citep[e.g.][]{1974A&AS...15..417H}, which favors sparsity.
For each epoch and regularization type, we explore a grid of regularization hyperparameters, and choose the final value using the ``L-curve" approach \citep[e.g.][]{2017JOSAA..34..904T}.
We plot the regularizer value ($f_{r}$) against the reconstructed image reduced $\chi^2$.
This resembles an ``L" where images having relatively constant $\chi^2$ with $f_{r}$ are under-regularized, and images having rapidly-changing $\chi^2$ with $f_r$ are over-regularized.
The transition between under- and over-regularization (the elbow in the curve) gives the optimal hyperparameter value.

\begin{figure*}[ht]
\begin{center}
\begin{tabular}{c}
\includegraphics[width=0.8\textwidth]{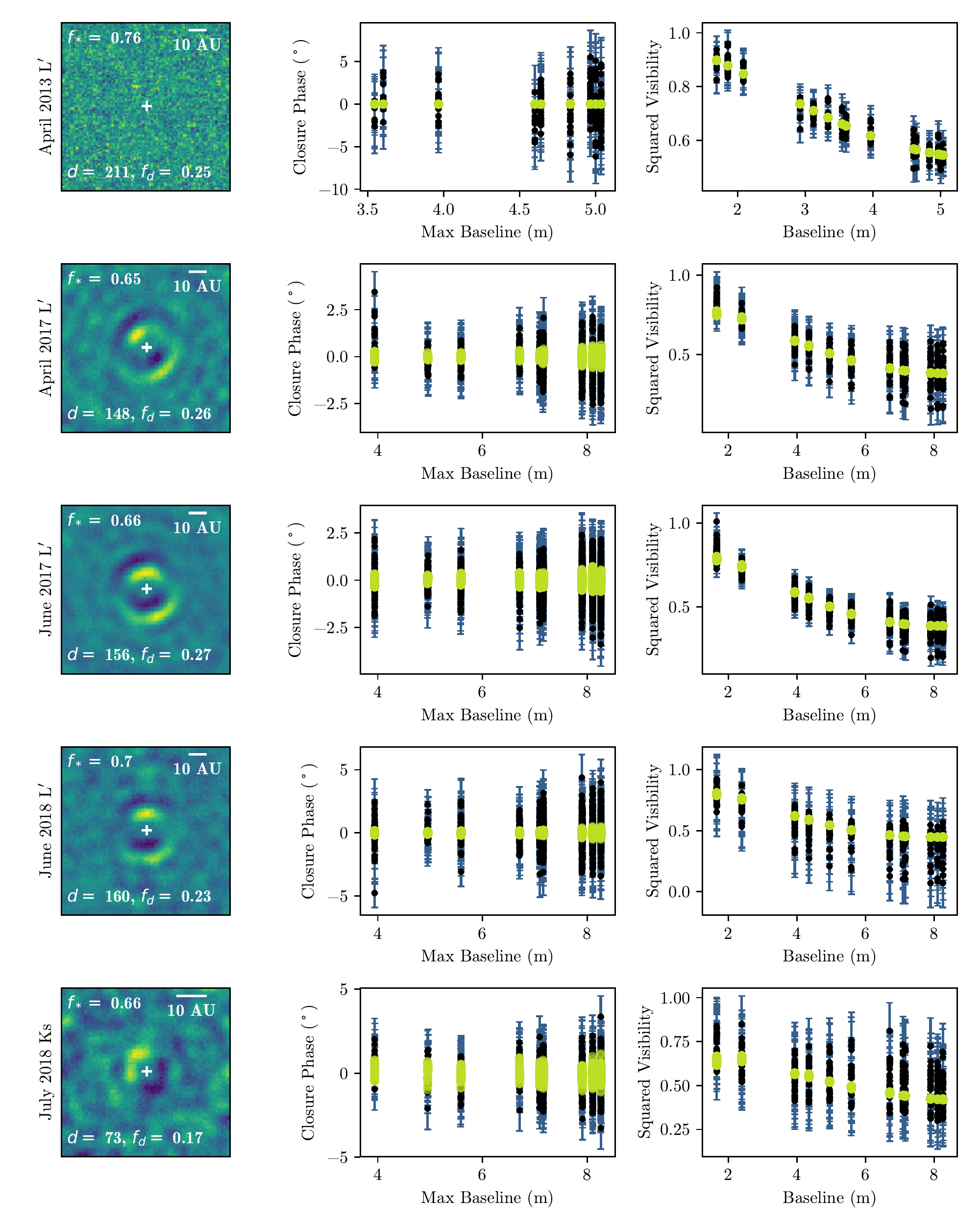}
\end{tabular}
\end{center}
\caption
{ \label{fig:recon-noreg}
Reconstructed images for each epoch (left), which include a central compact component with fractional flux $f_*$ and an extended uniform disk with diameter $d$ and fractional flux $f_d$. The center and right panels, respectively, show the observed closure phases and squared visibilities (black points with error bars), and the reconstructed image closure phases and squared visibilities (yellow points). White crosses mark the position of the central delta function. Images are shown north up east left.
}
\end{figure*}

Applying these optimal hyperparameter values showed that the choice of model components affected the reconstruction more significantly than the regularization choice. 
We thus show unregularized Bayesian images as the final reconstructed images in Figure \ref{fig:recon-noreg}.
For all L$'$ epochs, $65\%-70\%$ of the flux is contained in a central delta function component and $23\%-27\%$ in a symmetric disk component with a diameter of $\sim150$ mas.
The remaining flux is distributed asymmetrically, with the brightest peak to the northeast and a fainter peak to the southwest.
At K$\mathrm{_s}$, roughly the same amount of flux is contained in the compact component, and slightly less ($\sim17\%$) in a more compact ($\sim70$ mas) disk component.
The $\mathrm{K_s}$ images display an asymmetry to the east and northeast, but lack the pronounced asymmetry to the southwest.

We also explored image reconstructions with delta functions alone and resolved disks alone.
In these cases, SQUEEZE's ability to reproduce the kernel phases, and the observed asymmetric structure in the images, stayed the same. 
The algorithm could still reproduce the squared visibilities when a resolved disk (with no delta function) was included; in this case the preferred disk sizes were $\sim55$ mas at L$'$ and $\sim30$ mas at Ks.
However, SQUEEZE did not converge to an image that matched the squared visibilities when the only model component included was a central delta function. 
These tests suggest that the brightness distribution contains some level of resolved, symmetric structure.

\begin{figure*}[ht]
\begin{center}
\begin{tabular}{c}
\includegraphics[width=\textwidth]{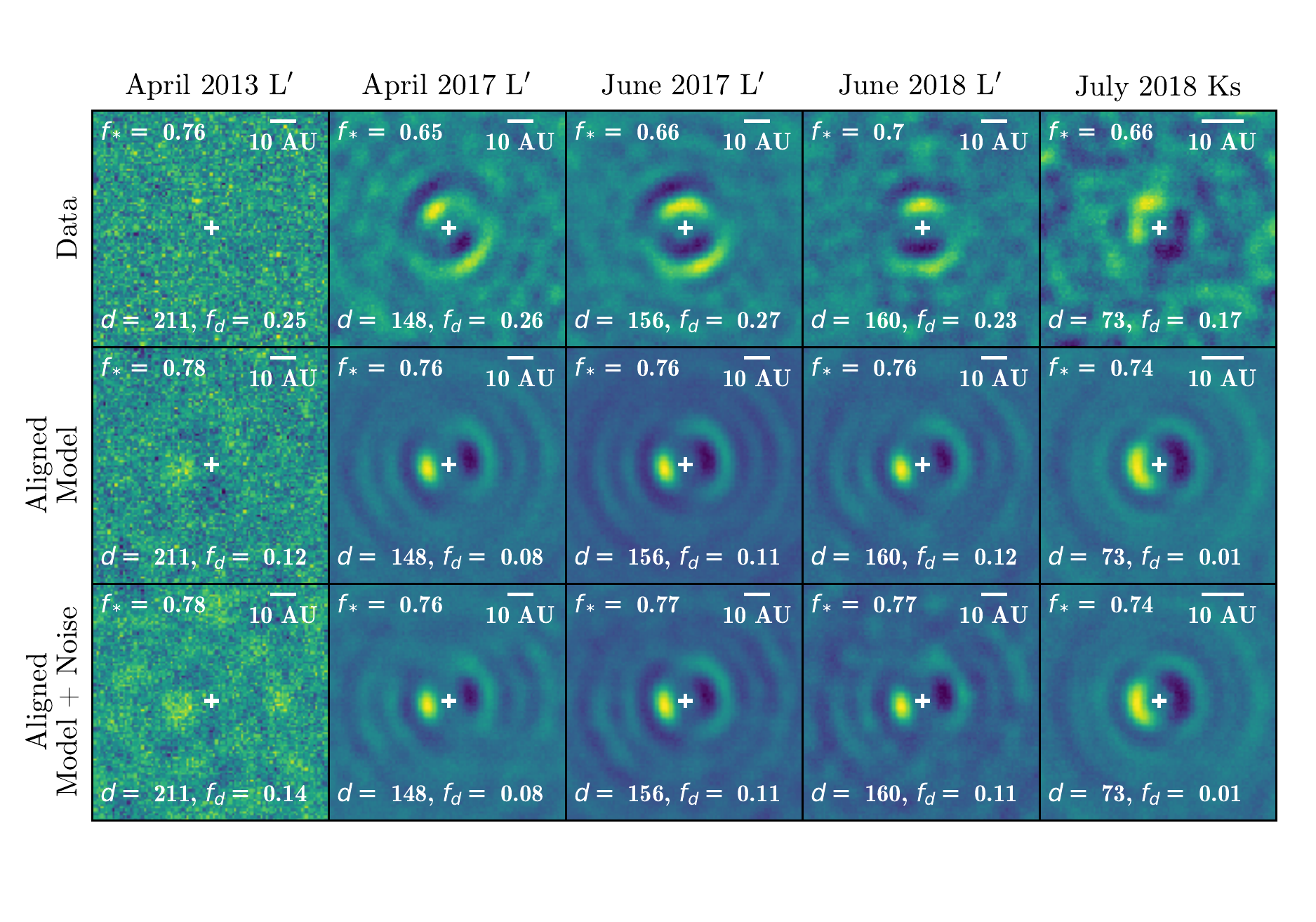}
\end{tabular}
\end{center}
\caption
{ \label{fig:recon-align}
Images reconstructed from the observations (top row), simulated observations of the aligned radiative transfer model shown in Figure \ref{fig:disk1} (middle row), and simulated observations of the aligned radiative transfer model plus noise (bottom row). For the Model + Noise simulations, we added Gaussian noise that would lead to squared visibility and closure phase histograms comparable to those in Figure \ref{fig:hists}. Each column shows a different observational epoch. White crosses mark the position of the central delta function. Images are shown north up east left.
}
\end{figure*}

\begin{figure*}[ht]
\begin{center}
\begin{tabular}{c}
\includegraphics[width=\textwidth]{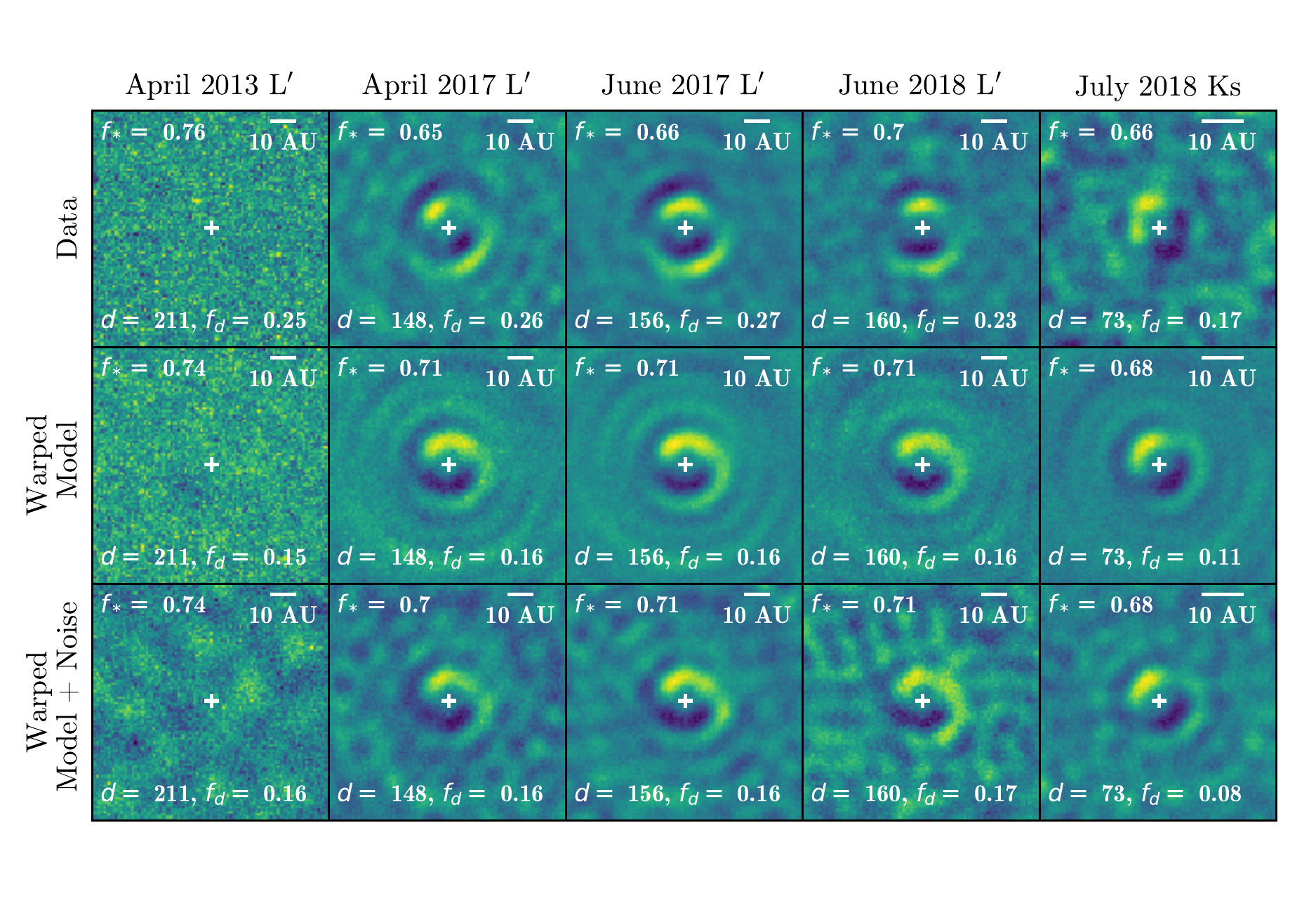}
\end{tabular}
\end{center}
\caption
{ \label{fig:recon-warp}
Images reconstructed from the observations (top row), simulated observations of the warped radiative transfer model shown in Figure \ref{fig:diskw} (middle row), and simulated observations of the warped radiative transfer model plus noise (bottom row). For the Model + Noise simulations, we added Gaussian noise that would lead to squared visibility and closure phase histograms comparable to those in Figure \ref{fig:hists}. Each column shows a different observational epoch. White crosses mark the position of the central delta function. Images are shown north up east left.
}
\end{figure*}

\begin{figure}[ht]
\begin{center}
\begin{tabular}{c}
\includegraphics[width=0.45\textwidth]{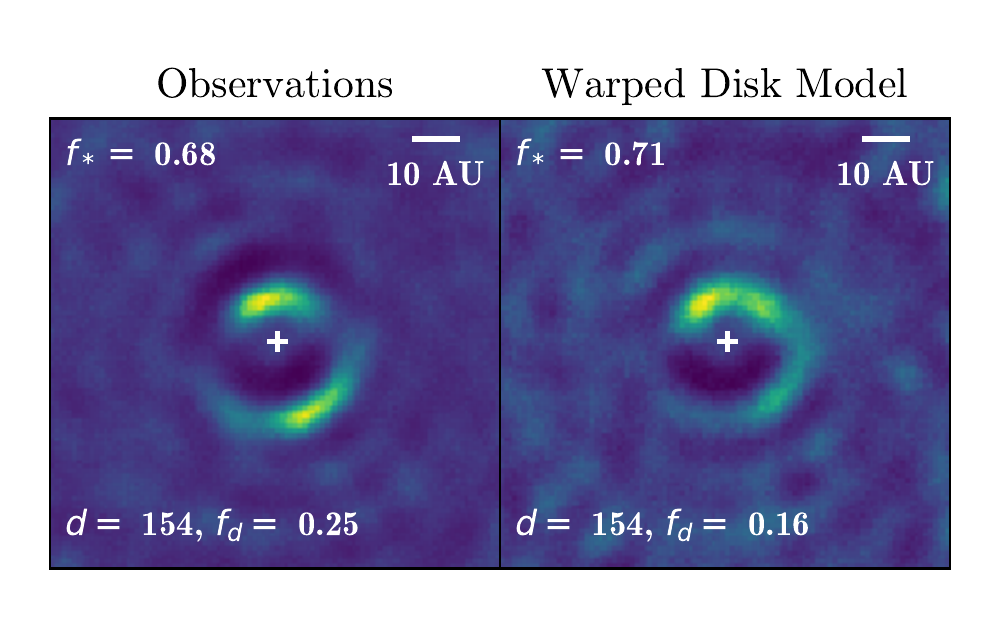}
\end{tabular}
\end{center}
\caption
{ \label{fig:recon-sim-lpc}
Images reconstructed from the combined L$'$ observations (left), and simulated multi-epoch observations of the disk model shown in Figure \ref{fig:diskw} with the same noise properties as the data (right). Images are shown north up east left.
}
\end{figure}

Lastly, we check whether any of the radiative transfer disk models can lead to reconstructed images similar to those presented in Figure \ref{fig:recon-noreg}.
For each epoch, we sample the radiative transfer models shown in Figures \ref{fig:disk1} and \ref{fig:diskw} with the same ($u,v$) coverage and sky rotation as the real data.
We reconstruct images before and after adding noise to the simulated observations so that their distributions match those shown in Figure \ref{fig:hists}.
We fix the size of the symmetric disk component to be the same as the best fit size for the real reconstructed images, and allow the amount of flux in the delta function and symmetric disk components to vary.
Figures \ref{fig:recon-align} - \ref{fig:recon-sim-lpc} show the results.

The aligned disk model cannot reproduce the reconstructed images: the peak position angle does not coincide with the peak position angle in the data reconstructions, and it is also constant between Ks and L$'$.
The warped disk model images roughly match the observations; they have an extended arc to the northeast that, with noise added, can have a peak position angle that changes slightly from epoch to epoch (middle versus bottom row in Figure \ref{fig:recon-warp}). 
This model can also reproduce the peak position angle shifts from Ks to L$'$, and provides a better match to the unresolved flux, although it slightly under-predicts the amount of flux in the extended symmetric component.
Among the simple disk models explored here, the models that could better reproduce the SQUEEZE symmetric disk component either had a large 10 $\mu$m excess in the spectral energy distribution, or were unable to reproduce the asymmetric structure in the reconstructed images.

\section{Discussion}\label{sec:disc}

\subsection{The SR 21 Small Grain Disk}
Previous observations of SR 21 require a puffy small-grain disk component, in addition to an optically thick small-grain disk, to reproduce the radial polarized intensity profiles at H band \citep{2013ApJ...767...10F}.
The data presented here also require a puffy disk component - all of the disk models that can reproduce the imaging and the spectral energy distribution require a large flaring index ($\beta\sim2$) and have typical scale heights of $\sim0.05-0.2$ AU at 1 AU for the small grains.
The disk model shown in Figure \ref{fig:diskw} provides a good match to the radial brightness profile presented in \citet{2013ApJ...767...10F}; it has roughly an r$^{-3}$ profile, with a discontinuity at 36 AU that is significantly smaller than the errors on the H band radial brightness profile. 
Although we use a simpler, two-component model, our imaging data support the hypothesis that SR 21 has a small-grain ``atmosphere" that may represent a disk wind or a remnant envelope.

Previous scattered light observations could not distinguish between a continuous small-grain disk down to separations of $\sim0.07$ AU (the sublimation radius), or a small-grain disk wall at $\sim7$ AU \citep{2013ApJ...767...10F}, where a gas disk truncation has been reported \citep{2008ApJ...684.1323P}.
The higher resolution imaging data presented here suggest a small-grain disk truncation at $\sim4-7$ AU.
This density distribution may be caused by the gravitational influence of a giant planet interior to the dust disk truncation, a scenario that has previously been modeled for SR 21 \citep[e.g.][]{2015A&A...573A...9P}.
This scenario would also be consistent with the observed gas disk truncation at $\sim7$ AU \citep{2008ApJ...684.1323P} as well as the gas density drop within the millimeter clearing \citep[e.g.][]{2016A&A...585A..58V}.
The large grain sizes required to match the imaging and SED also agree with this scenario, as grain growth can take place in pressure maxima induced by companions.

Reproducing the imaging data presented here requires an inner disk warp or spiral structure. 
The radiative transfer modeling and image reconstruction simulations in Figures \ref{fig:diskw} and \ref{fig:recon-warp} demonstrate this; matching the data requires a significant position angle change between the inner and outer regions of the disk. 
Position angle changes with semi-major axis have previously been inferred from SR 21's isophotes at larger radii \citep[][]{2013ApJ...767...10F}, and were thought to indicate spiral structure.
While a warped inner disk provides a reasonably good fit to the observations, similar results could likely be achieved in radiative transfer simulations of spiral structure.
In this case the Ks band reconstructions, which probe tighter angular separations than the L$'$ observations, would still exhibit an offset in peak position angle.
Disk structures such as warps and spirals have recently been reported in studies of other transition disks \citep[e.g.][]{2017AJ....153..264F,2018MNRAS.477.5104C,2019ApJ...872..122M}, and features like this can be caused by giant planet companions \citep[e.g.][]{2015ApJ...809L...5D}.\\

\subsection{Testing the Companion Scenario}
Previous mid-infrared observations of SR 21 suggested the presence of a companion near the edge of the inner disk clearing \citep{2009ApJ...698L.169E}.
This was interpreted as circumplanetary material around a forming sub-stellar object.
While the dominant features in the imaging data presented here are well explained by a small-grain disk component, the disk morphology suggests the influence of a massive companion.
Furthermore, the disk models explored here cannot reproduce SR 21's near-infrared excess (see Figures \ref{fig:disk1} and \ref{fig:diskw}) while matching the imaging and without over-predicting the $10~\mu$m silicate feature in the SED.
While this may be caused by the assumed dust composition or the simplicity of the disk model, circumplanetary material could account for the excess flux at these wavelengths. 

\begin{figure}[ht]
\begin{center}
\begin{tabular}{c} 
\includegraphics[width=0.47\textwidth]{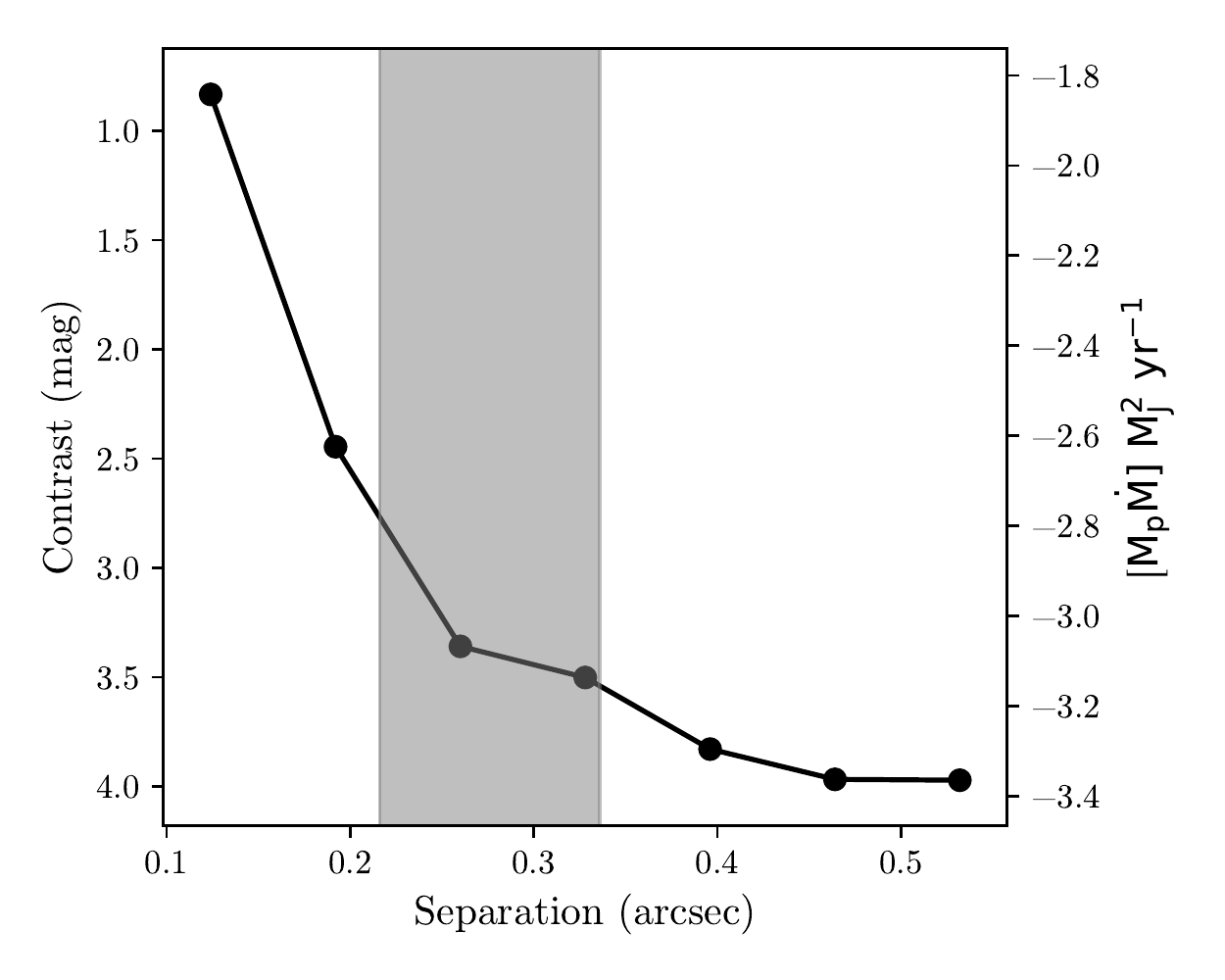}
\end{tabular}
\end{center}
\caption
{ \label{fig:visaocc}
Measured contrast curve for MagAO SDI observations. The grey shading indicates the 27-42 pixel region in the reduced images, corresponding to the AO control radius. The left y-axis shows contrast in magnitudes, and right y-axis converts that contrast to a detectable planet mass times accretion rate, assuming similar extinction to any companion as that for SR 21, and a planet radius of 1.6 $\mathrm{R_J}$.
}
\end{figure}

We generate a contrast curve from the H$\alpha$ SDI imaging data to check whether we can rule out this scenario (see Figure \ref{fig:visaocc}).
To estimate the contrast we first split the image into several 17-pixel (1 FWHM) wide annuli.
For each annulus, we low-pass filter by a Gaussian with the same FWHM as the stellar PSF to smooth out high frequency noise.
We then calculate the mean ($x$) and standard deviation ($s$) of each annulus, as well as the number of resolution elements ($n$) within the annulus.
Following \citet{Mawet:2014}, we assume a Student-t distribution with $n-1$ degrees of freedom and standard deviation $s$, and find the flux that would yield a Gaussian 5$\sigma$ false positive probability ($\sim3\times10^{-7}$).
We then multiply this value by $\sqrt{1+1/n}$ and add $x$ to calculate the $5\sigma$ contrast for a given separation / annulus.

We calibrate this contrast curve by injecting fake planets into the raw H$\alpha$ images.
We inject single companions at separations equal to each separation in the contrast curve, at 8 different position angles spaced by $45^\circ$.
We process each planet injection with identical KLIP parameters as the final SDI image, and measure the recovered planet flux.
For each fake planet, the throughput is the ratio of the recovered to input planet fluxes.
We take the mean throughput for the set of 8 fake planets as the throughput for a given annulus / separation.
We then divide each flux value in the contrast curve by its throughput to calibrate.

Figure \ref{fig:visaocc} (left y-axis) shows the resulting contrast curve.
We convert this achievable contrast to a detectable planet mass times accretion rate to place limits on any H$\alpha$-bright accreting companions. 
We assume $\mathrm{A_v} \sim 6.3$ for extinction to SR 21 \citep{2011ApJ...732...42A,2013ApJ...767...10F}:, and similar extinction toward the star as any hypothetical companions.
We also assume the same H$\alpha$ line luminosity to accretion luminosity scaling relations as those found for young stars \citep[e.g.][]{2012A&A...548A..56R}, and a planet radius of 1.6 $\mathrm{R_J}$.
The right y-axis of Figure \ref{fig:visaocc} shows the detectable planet masses times accretion rates as a function of separation.
We can only rule out accretion rates of $\sim10^{-2} - 10^{-3}$ $\mathrm{M_J^2~yr^{-1}}$.

We also check whether the $\mathrm{K_s}$ and L$'$ imaging data can constrain this scenario by exploring a small range of disk plus companion models.
We inject companions into the radiative transfer disk model shown in Figure \ref{fig:diskw}, varying companion semi-major axis, and $\mathrm{K_s}$ and L$'$ contrast with respect to the star.
We force the companion position to change from epoch to epoch according to a Keplerian orbit in the plane of the outer disk.
We thus also let the true anomaly vary.

A wide range of companion contrasts and semi-major axes are indistinguishable from the null (disk-only) model at $1\sigma$ using a $\chi^2$ interval. 
At L$'$, contrasts ranging from 1.5 to 6 magnitudes for semi-major axes of 1 - 5 AU are consistent with the disk-only model.
At $\mathrm{K_s}$, companions with contrasts of $\sim3-6$ magnitudes are indistinguishable from the null model for separations larger than $\sim2-3$ AU, and contrasts of $\sim0.5 - 3$ magnitudes at smaller separations.
For comparison, the companion model described in \citet{2009ApJ...698L.169E}, which had a temperature of 730 K and a radius of 40 R$_\odot$, had expected contrasts of 1.5 and 3.3 magnitudes at L$'$ and $\mathrm{K_s}$.
Thus we cannot rule out the companion parameters presented in \citet{2009ApJ...698L.169E}.

Many disk plus companion models provide significantly better fits to the Keck data than the disk-only model, with the best fit having contrasts of 3.5 and 6.0 at L$'$ and $\mathrm{K_s}$, respectively. 
Companions consistent with these contrasts are capable of filling in SR 21's near infrared excess from $\sim2-5~\mu$m.
Given the small number of disk and disk plus companion models explored, and the relatively low achievable L$'$ contrast within 5 AU, confirming or ruling out the companion scenario may require higher resolution data and more rigorous modeling efforts.

\section{Conclusions}\label{sec:conc}
We presented new, spatially resolved observations of the SR 21 transition disk in the visible and at $\mathrm{K_s}$ and L$'$.
The multi-epoch data are more consistent with forward scattering from a circumstellar disk than with an orbiting companion.
The reconstructed images reveal complex disk structure beyond a simple inclined rim, such as a warp or spiral.
The imaging also suggests an inner disk truncation at a few AU, and requires a puffy small grain component, in agreement with previous scattered light studies at shorter wavelengths.
The disk models explored here require relatively large grains ($\gtrsim2-5~\mu$m) to match both the imaging data and the spectral energy distribution.

The radiative transfer modeling and the reconstructed images support the hypothesis that SR 21 may be shaped by a giant planet companion.
This agrees with previous mid-infrared observations that suggested the presence of a $\sim700$ K companion, possibly corresponding to circumplanetary material around a forming sub-stellar object.
Our data allow for similar companion contrasts as those reported in \citet{2009ApJ...698L.169E}, and disk plus companion models with higher companion contrasts can provide a better fit to the data than disk models alone.
Future observations at multiple wavelengths and higher spatial resolution will evince SR 21's disk structure in greater detail, and will place tighter constraints on substellar companions in this system.

\acknowledgements
S.S. is supported by an NSF Astronomy and Astrophysics Postdoctoral Fellowship under award AST-1701489.
J.A.E. acknowledges support from NSF award number 1745406.
K.M.M.'s and L.M.C.'s work is supported by the NASA Exoplanets Research Program (XRP) by cooperative agreement NNX16AD44G.

The data presented herein were obtained at the W. M. Keck Observatory, which is operated as a scientific partnership among the California Institute of Technology, the University of California and the National Aeronautics and Space Administration. The Observatory was made possible by the generous financial support of the W. M. Keck Foundation.

The authors wish to recognize and acknowledge the very significant cultural role and reverence that the summit of Maunakea has always had within the indigenous Hawaiian community.  We are most fortunate to have the opportunity to conduct observations from this mountain.

\bibliography{/Users/stephaniesallum/Dropbox/PaperWritingTools/Latex/references2}

\end{document}